\documentclass[pra,twocolumn,citeautoscript,superscriptaddress]{revtex4}

\usepackage{url}
\usepackage{hyperref}
\usepackage{grffile}
\usepackage{graphicx}
\usepackage{amsmath}
\usepackage{amssymb}
\usepackage{cases}
\usepackage{xcolor}
\usepackage{mathrsfs}
\usepackage{braket}
\usepackage{multirow}
\usepackage[normalem]{ulem}
\renewcommand{\vec}{\mathbf}

\begin{document}

\title{Nonlinear transport in the presence of a local dissipation}

\author{A.-M. Visuri}
\email{avisuri@uni-bonn.de}
\affiliation{Physikalisches Institut, University of Bonn, Nussallee 12, 53115 Bonn, Germany}
\author{T. Giamarchi}
\affiliation{Department of Quantum Matter Physics, University of Geneva, 24 quai Ernest-Ansermet, 1211 Geneva, Switzerland}
\author{C. Kollath}
\affiliation{Physikalisches Institut, University of Bonn, Nussallee 12, 53115 Bonn, Germany}

\begin{abstract}
  We characterize the particle transport, particle loss, and nonequilibrium steady states in a dissipative one-dimensional lattice connected to reservoirs at both ends. The free-fermion reservoirs are fixed at different chemical potentials, giving rise to particle transport. The dissipation is due to a local particle loss acting on the center site. We compute the conserved current and loss current as functions of voltage in the nonlinear regime using a Keldysh description. The currents show step-like features which are affected differently by the local loss: The steps are either smoothened, nearly unaffected, or even enhanced,
  depending on the spatial symmetry of the single-particle eigenstate giving rise to the step.
Additionally, we compute the particle density and momentum distributions in the chain. At a finite voltage, two Fermi momenta can occur, connected to different wavelengths of Friedel oscillations on either side of the lossy site. We find that the wavelengths are determined by the chemical potentials in the reservoirs rather than the average density in the lattice.
\end{abstract}

\maketitle

\section{Introduction}

Understanding the role of dissipation is one of the most important questions in quantum physics, since dissipation can hardly be avoided in any physical system. The dissipative coupling of a quantum system to an environment generally leads to the exchange of energy and to quantum decoherence~\cite{BreuerPetruccione2002}. It is therefore often detrimental to applications taking advantage of quantum coherence. Controlled dissipation can, however, be an essential tool in the preparation and stabilization of novel, nonequilibrium quantum states~\cite{MuellerZoller2012,HarringtonMurch2022}, or the study of dissipative phase transitions~\cite{BenaryOtt2022}.
Examples include the preparation of squeezed states with ultracold atoms~\cite{CaballarWatanabe2014}, a Tonks-Girardeau gas of molecules~\cite{SyassenDuerr2008}, or entanglement among trapped ions~\cite{BarreiroBlatt2011}, and
the dissipative stabilization of a photon Mott insulator~\cite{MaSchuster2019}.
Dissipation engineering can also be used as a tool in quantum information processing~\cite{VerstraeteCirac2009,HarringtonMurch2022} and to control quantum transport~\cite{DamanetDaley2019,DamanetDaley2019Nov}. 

In the recent years, a new experimental platform has emerged to study the effects of dissipation. Cold atom experiments allow to almost perfectly isolate quantum systems from their environment, but also to engineer dissipation processes in a controlled way, for example in the form of local particle losses~\cite{BarontiniOtt2013,LabouvieOtt2016,CormanEsslinger2019,LebratEsslinger2019,BenaryOtt2022}.
Theoretically, local losses and dephasing have been investigated in weakly-interacting~\cite{BrazhnyiOtt2009,TonielliMarino2020,WillFleischhauer2022} and hard-core~\cite{DuttaCooper2020,DuttaCooper2021} bosonic atoms, the Bose-Hubbard model~\cite{BarmettlerKollath2011,WitthautWimberger2011,KieferSirker2017,RossiniMazza2021}, and fermions in one~\cite{WolffKollath2020,FromlDiehl2019,FromlDiehl2020,MullerDiehl2021,AlbaCarollo2022} and two~\cite{WasakPiazza2021} dimensions, with focus on the presence of the quantum Zeno effect~\cite{MisraSudarshan1977,BreuerPetruccione2002}. Further recent studies have elucidated the effect of dissipation and dephasing on transport properties.

Steady-state transport through a system, e.g. wire or quantum dot, coupled to leads is a nonequilibrium situation with much practical importance in nanotechnology~\cite{NazarovBlanter2009,Ryndyk2015}, and it is one of the most common ways to characterize the properties of new materials or devices. 
Whereas transport experiments were initially mostly used for probing solid state devices, transport setups have recently also been engineered with cold atoms. For example, cold-atom analogues of a two-terminal transport measurement~\cite{KrinnerBrantut2017} realizes a typical setup of mesoscopic devices, where the system of interest is coupled to two leads at different chemical potentials. In these cold-atom experiments, it is now possible to investigate the effects of dissipation on particle transport in a controlled way. In particular, such experiments offer a possibility to study the effects of particle losses or dephasing on a nonequilibrium steady state, generated by a chemical potential difference.
Theoretically, it has been shown that dephasing can lead to diffusive transport in quantum coherent systems, while in the presence of disorder, delocalization and noise-assisted transport can arise from dephasing (see Ref.~\cite{LandiSchaller2022} and references therein). In the case of local particle losses, it was shown that transport through a one-dimensional lattice can be robust to a local loss~\cite{VisuriKollath2022}. On the other hand, a cold-atom setup with a lossy quantum point contact was used to demonstrate a reduction of conductance plateaus~\cite{CormanEsslinger2019} and a robustness of superfluid transport to particle loss~\cite{HuangEsslinger2022}.

The theoretical treatment of transport can be performed via different routes. While in the linear-response regime, where the external field is small, conductivity is generally given by the Kubo formula~\cite{Mahan2000}, at finite voltages, other theoretical methods are required. In the case of noninteracting fermions, coherent transport can be described in the Landauer-B\"uttiker formalism~\cite{Landauer1957}. 
For interacting particles, in the nonlinear regime, most theoretical descriptions are based on nonequilibrium Green's function methods~\cite{MeirWingreen1992,Rammer2007,HaugJauho2008} where the system coupled to leads is described by a Hamiltonian operator and the time evolution is unitary, or quantum master equations~\cite{LandiSchaller2022,MaksimovKolovsky2022}, where the coupling to the leads is modeled by particle losses and gains at the boundaries. Quantum master equation approaches are valid when the system-reservoir coupling is weak, while nonequilibrium Green's function techniques can be applied at any coupling strength but interactions are typically taken into account only approximately.
The analytic correspondence between the Hamiltonian evolution of a system coupled to fermionic (or bosonic) reservoirs and the Lindblad evolution of an open quantum system with losses and gains at the boundaries is an interesting question and has inspired recent theoretical studies~\cite{JinGiamarchi2020, Uchino2022}.

Here, we study theoretically a local particle loss in a one-dimensional lattice coupled to fermionic reservoirs. An approximate way to model systems of this type is a non-Hermitian Hamiltonian within the Landauer-B\"uttiker formalism~\cite{CormanEsslinger2019}, but an exact solution can be found through nonequilibrium Green's functions written in the Keldysh formalism~\cite{KamenevBook} extended to open quantum systems~\cite{SiebererDiehl2016}. The lossy site in our model is governed by Lindblad evolution, while the other lattice sites and the reservoirs evolve unitarily.
This system's conductance, which measures transport in the zero-voltage limit, was analyzed in detail in our previous work~\cite{VisuriKollath2022}, and we focus here on the finite-voltage regime. We explore the effects of a local dissipation on the nonlinear current-voltage characteristics, which in the absence of dissipation have step-like features. A step-like voltage dependence is also found for the loss current. Interestingly, the analytic form of the currents coincides with a system where the particle loss is replaced by a third terminal, given that certain conditions, such as the absence of gain from the third terminal, are satisfied~\cite{Uchino2022}.
Furthermore, we analyze the loss current and the momentum and density distributions in the lattice.
The momentum distribution shows the presence of two Fermi surfaces, with Fermi momenta determined by the chemical potentials of the reservoirs. This is reflected in the Friedel oscillations, the wavevector of which changes across the lossy site.

The paper is organized as follows: The model for the open quantum system is introduced in Sec.~\ref{sec:model}. Section~\ref{sec:model} also introduces the relevant quantities to characterize transport, particle loss, and the nonequilibrium steady states, and summarizes the calculation of nonequilibrium correlation functions in the Keldysh formalism. This section contains and expands some of the points discussed in Ref.~\cite{VisuriKollath2022}. Before discussing the results, a simple equilibrium model is presented in Sec.~\ref{sec:equilibrium} to gain understanding of certain features of the nonequilibrium observables. In Sections~\ref{sec:current-voltage}--\ref{sec:imbalance}, we analyze the current-voltage characteristics and loss current and discuss properties of the steady states.
Conclusions and an outlook are given in Sec.~\ref{sec:conclusions}.
Results for additional parameters as well as technical details are presented in the Appendices.

\section{Model and methods}
\label{sec:model}

\subsection{Quantum master equation}

\begin{figure}
\includegraphics[width=\linewidth]{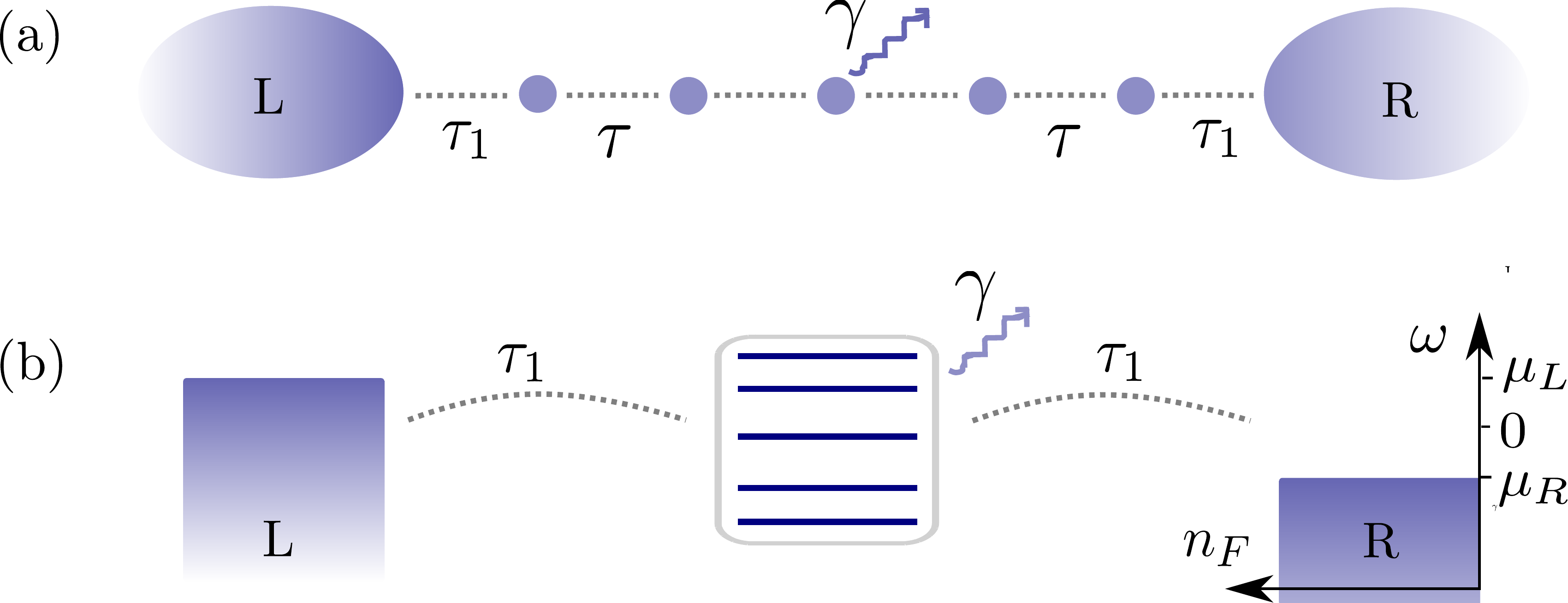}
\caption{(a) A lattice of $M$ sites is connected to reservoirs at both ends and a local particle loss with amplitude $\gamma$ acts on the center site. The coupling to the reservoirs $\tau_1$ is in general different from the tunneling amplitude $\tau$ within the lattice. (b) Representation of the different energy scales. In the reservoirs, states up to the chemical potentials $\mu_{L, R}$ are filled at zero temperature, as shown by the Fermi distributions $n_F(\omega)$ on either side. The eigenenergies of an isolated lattice are depicted by the horizontal lines within the box in the middle. The current through the lattice changes in steps when the chemical potentials coincide with eigenenergies of the lattice. Adapted from Ref.~\cite{VisuriKollath2022}.}
\label{fig:dissipative_chain}
\end{figure}

The system is depicted in Fig.~\ref{fig:dissipative_chain}: a one-dimensional lattice is coupled at both sides to a free-fermion reservoir and subjected to a local particle loss acting on the central site. 
In the absence of loss, the system is described by the Hamiltonian
\begin{equation}
H = \sum_{i = L, R} H_i + H_{\text{chain}} + H_{\text{t}},
\label{eq:hamiltonian}
\end{equation}
where the indices $L$ and $R$ denote the left and right reservoirs, respectively. The reservoirs are described by the free-fermion Hamiltonian  
\begin{equation}
H_i = \sum_{\mathbf{k}} \left(\epsilon_{\mathbf{k}} - \mu_i \right) \psi_{i \mathbf{k}}^{\dagger} \psi_{i \mathbf{k}}^{\phantom{\dagger}},
\label{eq:hamiltonian_leads}
\end{equation}
where $\psi_{i \mathbf{k}}^{\dagger}$ ($\psi_{i \mathbf{k}}^{\phantom{\dagger}}$) is the fermionic creation (annihilation) operator acting on reservoir $i$, $\mathbf{k}$ denotes momentum, and $\epsilon_{\mathbf{k}}$ is the energy. We set $\hbar = 1$ for simplicity. The chemical potential $\mu_i$ of the reservoirs is in general different for $i = L, R$, which imposes a voltage $V = \mu_L - \mu_R$ between the reservoirs. In the following, we choose the chemical potentials symmetrically as $\mu_{L, R} = \pm V/2$.
We assume that the density of states of the reservoirs is a constant and thus have the linear dispersion relation $\epsilon_{\mathbf{k}} = v_F \left(\mathbf{k} - \mathbf{k}_F \right)$, where $v_F$ is the Fermi velocity and $\mathbf{k}_F$ the Fermi momentum.

The Hamiltonian operator for the lattice is
\begin{equation}
H_{\text{chain}} = \epsilon \sum_{j = -l}^l d_j^{\dagger} d_j -\tau \sum_{j = -l}^{l - 1} \left( d_{j + 1}^{\dagger} d_j + \text{H.c.} \right),
\label{eq:chain_hamiltonian}
\end{equation}
where $d_j^{\dagger}$ ($d_j$) is the fermionic creation (annihilation) operator acting on site $j$ and $\tau$ is the tunneling amplitude within the chain. The lattice spacing is set to 1. We also consider the case of a single site, or quantum dot, and the tunneling term in Eq.~(\ref{eq:chain_hamiltonian}) only exists if the chain has more than one site. Additionally, we consider an energy offset $\epsilon$ which is equal for all sites in the chain. The length of the chain is $M = 2 l + 1$, so that the results presented here are for odd $M$. The value $M=1$ corresponds to a quantum dot. 
The Hamiltonian
\begin{equation}
H_{\text{t}} = -\tau_1 \left[\psi^{\dagger}_{L}(\mathbf{0}) d_{-l} + d_l^{\dagger} \psi^{\phantom{\dagger}}_{R}(\mathbf{0}) + \text{H.c.}\right],
\label{eq:hamiltonian_tun}
\end{equation}
written in position basis, describes the tunneling between the ends of the chain and the respective reservoirs. In momentum basis, the field operators at $\vec{r} = \vec{0}$ are given by $\psi^{\dagger}_i(\mathbf{0}) = \sum_{\vec{k}} \psi_{i \vec{k}}^{\dagger}$. The tunneling occurs at one spatial point $\vec{r} = \vec{0}$ in each reservoir, with tunneling amplitude~$\tau_1$.

In the presence of the particle loss on the center site, we use the quantum master equation
\begin{equation}
\frac{d\rho}{dt} = -i [H, \rho] + \gamma \left[ L \rho L^{\dagger} - \frac{1}{2} \left\{ L^{\dagger} L, \rho \right\} \right],
\label{eq:master_equation}
\end{equation}
which gives the time evolution of the density operator~$\rho$. The loss rate is denoted by $\gamma$, and the Lindblad jump operator $L$ is here the annihilation operator at site $j = 0$, i.e.~$L = d_0$, representing the losses. In the following, numerical values of the parameters are reported as dimensionless, so that $V$, $\epsilon$, and $\gamma$ are in units of the lattice tunneling amplitude $\tau$, the lattice-reservoir coupling $\tau_1$ is in units of $\tau \sqrt{\mathcal{V}}$, where $\mathcal{V}$ is the volume of the reservoirs, and $\tau$ is fixed to $\tau = 1/(\pi \rho_0 \mathcal{V})$, where $\rho_0$ is the constant density of states of the reservoirs per unit volume. We additionally define the parameter $\Gamma = \pi \rho_0 \tau_1^2$ with units of energy.

\subsection{Observables}

To characterize transport, we calculate the conserved particle current $I$ through the lattice system. The current is connected to the change of particle numbers in the reservoirs, 
\begin{align}
I &= -\frac{1}{2} \frac{d}{dt}\braket{N_L - N_R} \label{eq:conserved_current}
\\
\begin{split}
&= -\frac{i \tau_1}{2} \Big(\braket{d_{-l}^{\dagger} \psi_L(\mathbf{0})} - \braket{\psi_L^{\dagger}(\mathbf{0}) d_{-l}} \\
&\hspace{2cm}+ \braket{\psi_R^{\dagger}(\mathbf{0}) d_{l}} - \braket{d_{l}^{\dagger} \psi_R(\mathbf{0})} \Big), 
\end{split} \label{eq:operator_form}
\end{align}
where the second and third lines are obtained through Eq.~(\ref{eq:master_equation}) (see Refs.~\cite{VisuriKollath2022,JinGiamarchi2020,Uchino2022} for details). The expectation values are defined as $\braket{A} = \text{Tr}(A \rho)$ for a generic operator~$A$, and the particle number operator is $N_i = \int d\mathbf{r} \psi_i^{\dagger}(\mathbf{r}) \psi_i^{\phantom{\dagger}}(\mathbf{r})$ with $i = L, R$.
We also compute the loss current
\begin{equation}
I_{\text{loss}} = -\frac{d}{dt}\braket{N_L + N_R} = \gamma \braket{n_0},
\label{eq:loss_current}
\end{equation}
which is connected to the particle number at the center site, $\braket{n_0} = \braket{d_0^{\dagger} d_0}$ (see Appendix~\ref{app:loss_current}).

In order to characterize the effect of the local dissipation on the nonequilibrium steady states, we calculate the particle density $\braket{n_j} = \braket{d_j^{\dagger} d_j}$ and momentum distribution in the lattice,
\begin{equation}
\braket{n_k} = \braket{d_k^{\dagger} d_k^{\phantom{\dagger}}} = \sum_{i, j = 1}^{M} \varphi_{i, k} \varphi_{j, k} \braket{d_i^{\dagger} d_j^{\phantom{\dagger}}},
\label{eq:momentum_distribution}
\end{equation}
where the indexing of the lattice sites is shifted to $i, j \in \{1, ..., M\}$ for simplicity. We use the basis functions of an isolated lattice, not coupled to leads, with open boundary conditions,
\begin{equation*}
\varphi_{j, k} = \sqrt{\frac{2}{M + 1}} \sin(k j).
\end{equation*}
The quasimomentum has the discrete values $k = n \pi/(M + 1)$ with $n \in \{1, 2, ..., M\}$.

\subsection{Keldysh formalism}

To compute nonequilibrium expectation values in the steady state, we use the functional integral formulation of the Keldysh formalism~\cite{KamenevBook}, where the integration extends over a closed time contour. The Keldysh action $S$ is written as a sum of the coherent and dissipative terms, 
\begin{equation}
S = \sum_{i = L, R} S_i + S_{\text{chain}} + S_{\tau_1} + S_{\text{loss}}.
\label{eq:action_sum}
\end{equation}
The first terms $S_i$ correspond to the reservoirs $i = L, R$, the second one represents the one-dimensional chain, and the third the coupling of the chain and the reservoirs. The local dissipation at the center site is described by the term $S_{\text{loss}}$.
The action is written in the basis of fermionic coherent states parametrized by the Grassmann variables $\psi = (\psi^+, \psi^-)$, where the vector elements correspond to the forward and backward time contours. 

We apply the bosonic convention to perform the Keldysh rotation into a basis $(\psi^1, \psi^2)$, where the action for the uncoupled reservoirs has the form 
\begin{equation}
\mathcal{S}_i = \int_{-\infty}^{\infty} \frac{d \omega}{2 \pi} 
\begin{pmatrix}
\bar{\psi}^1_i &\bar{\psi}^2_i
\end{pmatrix}
\mathcal{G}_i^{-1}(\omega)
\begin{pmatrix}
\psi_i^1	\\
\psi_i^2
\end{pmatrix}.
\label{eq:action_matrix}
\end{equation} 
In this basis, the inverse Green's function $\mathcal{G}_i^{-1}(\omega)$ has the standard matrix structure 
\begin{equation}
\mathcal{G}_i^{-1}(\omega) = 
\begin{pmatrix}
0	&\left[ \mathcal{G}_i^{\mathcal{A}} \right]^{-1} \\
\left[ \mathcal{G}_i^{\mathcal{R}} \right]^{-1}	&\left[ \mathcal{G}_i^{-1} \right]^{\mathcal{K}}
\end{pmatrix},
\label{eq:general_inverse_greens_function}
\end{equation}
where $\mathcal{G}_i^{\mathcal{A}}$, $\mathcal{G}_i^{\mathcal{R}}$, and $\mathcal{G}_i^{\mathcal{K}}$ are the advanced, retarded, and Keldysh Green's functions. 
Since the steady-state correlation functions do not depend on time, it is convenient to use the frequency representation. 
The Keldysh component is given by 
\begin{equation}
\mathcal{G}_i^{\mathcal{K}} = (\mathcal{G}_i^{\mathcal{R}} - \mathcal{G}_i^{\mathcal{A}}) [1 - 2 n_F(\omega - \mu_i)],
\label{eq:Keldysh_component}
\end{equation}
and $\left[ \mathcal{G}_i^{-1}\right]^{\mathcal{K}} = -\left[ \mathcal{G}_i^{\mathcal{R}} \right]^{-1} \mathcal{G}_i^{\mathcal{K}} \left[ \mathcal{G}_i^{\mathcal{A}} \right]^{-1}$. Here, $n_F(\omega) = (e^{\omega/T} + 1)^{-1}$ denotes the Fermi-Dirac distribution at temperature $T$. We set temperature to zero in both reservoirs and use natural units where $k_B = 1$. 

The reservoirs are modeled by local Green's functions at the point $\vec{r} = \vec{0}$ where the tunneling occurs,
\begin{align}
\mathcal{G}^{\mathcal{R}, \mathcal{A}}_{L/R}(\vec{r} = \vec{0}, \omega) = \frac{1}{\mathcal{V}} \sum_{|\mathbf{k}| \leq \Lambda/v_F} \frac{1}{\omega - \epsilon_{\mathbf{k}} \pm i \eta}.
\label{eq:greens_function_r0}
\end{align}
Here, $\mathcal{V}$ denotes the volume of the reservoirs, and $i\eta$ is an infinitesimal imaginary part. As the linear dispersion relation is unbounded, we set formally a cutoff $\pm \Lambda/v_F$ on the reservoir spectrum. We mostly focus on the limit $\Lambda \to \infty$, where the real part of Eq.~(\ref{eq:greens_function_r0}) vanishes. A finite cutoff and a finite real part of $\mathcal{G}^{\mathcal{R}, \mathcal{A}}_{L/R}$ is connected to the appearance of bound states outside the reservoir energy continuum, which we discuss in detail in Sec.~\ref{sec:equilibrium}.

The action for the one-dimensional chain consists of two contributions,
\begin{equation}
S_{\text{chain}} = \sum_{j = -l}^{l} S_j + \sum_{j = -l}^{l - 1} S_{j,\tau}.
\label{eq:chain_action}
\end{equation}
Here, $S_j$ is the action for the different lattice sites and has the same form as Eq.~(\ref{eq:action_matrix}) for the reservoirs, where the retarded and advanced Green's functions for the lattice sites are $\mathcal{G}_j^{\mathcal{R}, \mathcal{A}} = (\omega - \epsilon \pm i\eta)^{-1}$ and $\mathcal{G}_j^{\mathcal{K}}$ is given by Eq.~(\ref{eq:Keldysh_component}).
The second term in Eq.~(\ref{eq:chain_action}) corresponds to tunneling within the lattice. The loss term $S_{\text{loss}}$ is added to the action of the central site~\cite{SiebererDiehl2016},
\begin{equation*}
S_{\text{loss}} = 
\int \frac{d \omega}{2 \pi} 
\begin{pmatrix}
\bar{d}^1_0 &\bar{d}^2_0
\end{pmatrix}
\begin{pmatrix}
0			&-\frac{i\gamma}{2} \ \\
\frac{i\gamma}{2}	&i\gamma 
\end{pmatrix}
\begin{pmatrix}
d_0^1	\\
d_0^2
\end{pmatrix}.
\end{equation*}

The action $S$ for the full system can be written in matrix form as 
\begin{equation}
S=\int d \omega/(2 \pi) \bar{\mathbf{\Psi}}(\omega) \mathcal{G}^{-1}(\omega) \mathbf{\Psi}(\omega). 
\end{equation}
Here, we write the inverse Green's function $G^{-1}$ in the basis of all the different fields
\begin{equation}
\mathbf{\Psi} = 
\begin{pmatrix}
  \psi_L^1 &\psi_L^2	&d_{-l}^1	&d_{-l}^2
  &\dots&d_l^1	&d_l^2	&\psi_R^1 &\psi_R^2 
\end{pmatrix}^T
\label{eq:grassmann_basis}
\end{equation}
in a tridiagonal block form
\begin{align}
\mathcal{G}^{-1} = 
\begin{pmatrix}
L	&T_1	&0		&		&\dots		&		&0	\\
T_1	&D_{-l}	&T		&0	\\
0	&T		&\ddots	&T	\\
	&0		&T		&D_0	&T			\\
\vdots	&	&		&T		&\ddots	&T	\\
	&		&		&		&T		&D_l	&T_1	\\	
0	&		&		&		&		&T_1	&R
\end{pmatrix}
\label{eq:inverse_greens_function}
\end{align}
consisting of complex-valued matrix blocks of size $2 \times 2$. The corner blocks $L$ and $R$ correspond to the leads, which in the case of an unbounded reservoir spectrum read
\begin{equation}
L/R = 
\begin{pmatrix}
0			&-\frac{i}{\pi \rho_0} \\
\frac{i}{\pi \rho_0}	&\frac{2 i}{\pi \rho_0} \tanh \left( \frac{\omega - \epsilon}{2 T} \right)
  \end{pmatrix},
\end{equation}
where $\rho_0$ is the constant density of states per unit volume of the reservoirs. The matrix blocks corresponding to the lattice sites apart from $j = 0$ are
\begin{equation}
D_{j \neq 0} = 
\begin{pmatrix}
0			&\omega - \epsilon - i \eta \\
\omega - \epsilon + i \eta	&2 i \eta \tanh \left( \frac{\omega - \epsilon}{2 T} \right)
  \end{pmatrix},
\end{equation}
and the block for the central site is
\begin{equation}
D_{j = 0} =
  \begin{pmatrix}
0			&\omega - \epsilon - i \gamma/2 \\
\omega - \epsilon + i \gamma/2	& i \gamma
  \end{pmatrix}.
\end{equation}
For $D_{j=0}$, the infinitesimal imaginary term $i\eta$ is suppressed due to the finite imaginary part arising from the loss term. 
The tunneling matrix elements are contained in the off-diagonal blocks 
\begin{equation*}
T_1 = 
\begin{pmatrix}
0	&\tau_1	\\
\tau_1	&0
\end{pmatrix},
\hspace{1cm}
T = 
\begin{pmatrix}
0	&\tau	\\
\tau	&0
\end{pmatrix}.
\end{equation*}

The expectation value of the current $I$ of Eq.~(\ref{eq:conserved_current}) is written in terms of the Grassmann variables as
\begin{align}
\begin{split}
I &= \frac{i \tau_1}{4} \int_{-\infty}^{\infty}\frac{d \omega}{2 \pi} \Big( \braket{d_{-l}^1 \bar{\psi}_L^1(\mathbf{0})} - \braket{\psi_L^1(\mathbf{0}) \bar{d}^1_{-l}}  \\
&\hspace{3cm}+ \braket{\psi_R^1(\mathbf{0}) \bar{d}_l^1} - \braket{d_l^1 \bar{\psi}_R^1(\mathbf{0})} \Big) 
\end{split} \label{eq:current_grassmann}
\end{align} 
and the particle density in the lattice is given by
\begin{equation}
\braket{n_j} = \frac{1}{2} \int_{-\infty}^{\infty}\frac{d \omega}{2 \pi} \left( \braket{d_j^1 \bar{d}_j^1} - \braket{d_j^1 \bar{d}_j^2} + \braket{d_j^2 \bar{d}_j^1} \right).
\label{eq:density_grassmann}
\end{equation}
The matrix formulation of the action provides a simple algorithm for obtaining two-operator correlation functions by matrix inversion.
For the quadratic action of Eq.~(\ref{eq:action_sum}), correlation functions such as in Eqs.~(\ref{eq:current_grassmann}) and~(\ref{eq:density_grassmann}) are written as Gaussian path-integrals,
\begin{equation}
\braket{\psi^a \bar{\psi}^b} = \int \mathcal{D}[\bar{\psi}, \psi] \psi^a \bar{\psi}^{b} e^{i S[\bar{\psi}, \psi]} = i \mathcal{G}_{ab},
\label{eq:correlation_function}
\end{equation}
where $a, b$ denote the relevant indices in the basis~(\ref{eq:grassmann_basis}). Two-operator correlation functions can be obtained as the matrix elements of $\mathcal{G}$ simply by inverting $\mathcal{G}^{-1}$~\cite{BolechGiamarchi2004,BolechGiamarchi2005,HusmannBrantut2015,YaoZhai2018,VisuriKollath2022,HuangEsslinger2022}. The matrix inversion can be done analytically for small lattice sizes, and for large lattices, such as studied in Sec.~\ref{sec:friedel_oscillations} and~\ref{sec:imbalance}, it provides a convenient numerical algorithm. Note that this formulation is equivalent to solving the Keldysh Green's functions from the Dyson equation~\cite{MeirWingreen1992,Rammer2007,HaugJauho2008,Uchino2022}.
Further details can be found in Refs.~\cite{VisuriKollath2022,Uchino2022}.

\section{Current-voltage characteristics}
\label{sec:current-voltage}

A change in the particle number of the reservoirs is connected to particle currents. More precisely, a nonzero time derivative of the reservoir particle number results from two factors: The first is the flow of particles from one reservoir to the other when there is a chemical potential difference between the reservoirs. This is the conserved current defined by Eq.~(\ref{eq:conserved_current}). The second is the loss current due to the particle loss. It is nonzero also in the absence of a chemical potential difference and is given by Eq.~(\ref{eq:loss_current}). In this section, we focus on the conserved current, which is typically used to characterize transport, here through the lossy system. 
Evaluating the expectation values in Eq.~(\ref{eq:operator_form}) gives the expression
\begin{equation}
I = \int_{-\infty}^{\infty} \frac{d \omega}{2 \pi} g(\omega) \left[ n_L(\omega) - n_R(\omega) \right]
\label{eq:conserved_current_integral}
\end{equation}
for the conserved current. Here, we define $n_i(\omega) = n_F(\omega - \mu_i)$. In this section, we discuss the limit of an infinite energy continuum in the reservoirs $\Lambda \to \infty$. The consequences of a finite cutoff $\Lambda$ are detailed in Appendix~\ref{app:finite_continuum}.

\subsection{Quantum dot}

We first discuss a single lossy quantum dot coupled to leads. In this case, $g(\omega)$ has the form of a Lorentzian distribution
\begin{equation}
g(\omega) = \frac{4 \Gamma (\gamma + 4 \Gamma)}{(\gamma + 4 \Gamma)^2 + 4 \left(\omega - \epsilon \right)^2}.
\label{eq:quantum_dot_infinite_cutoff}
\end{equation}
The distribution is centered around the chemical potential of the dot $\epsilon$. 
Physically, the width of the distribution $\gamma + 4\Gamma$, where $\Gamma = \pi \rho_0 \tau_1^2$, corresponds to the inverse lifetime of the particle at the quantum dot. Both a larger tunneling $\tau_1$ between the quantum dot and the reservoirs and a larger loss rate $\gamma$ from the dot lead to a broadening of the distribution, connected to a shorter lifetime.

\begin{figure}[h!]
\includegraphics[width=\linewidth]{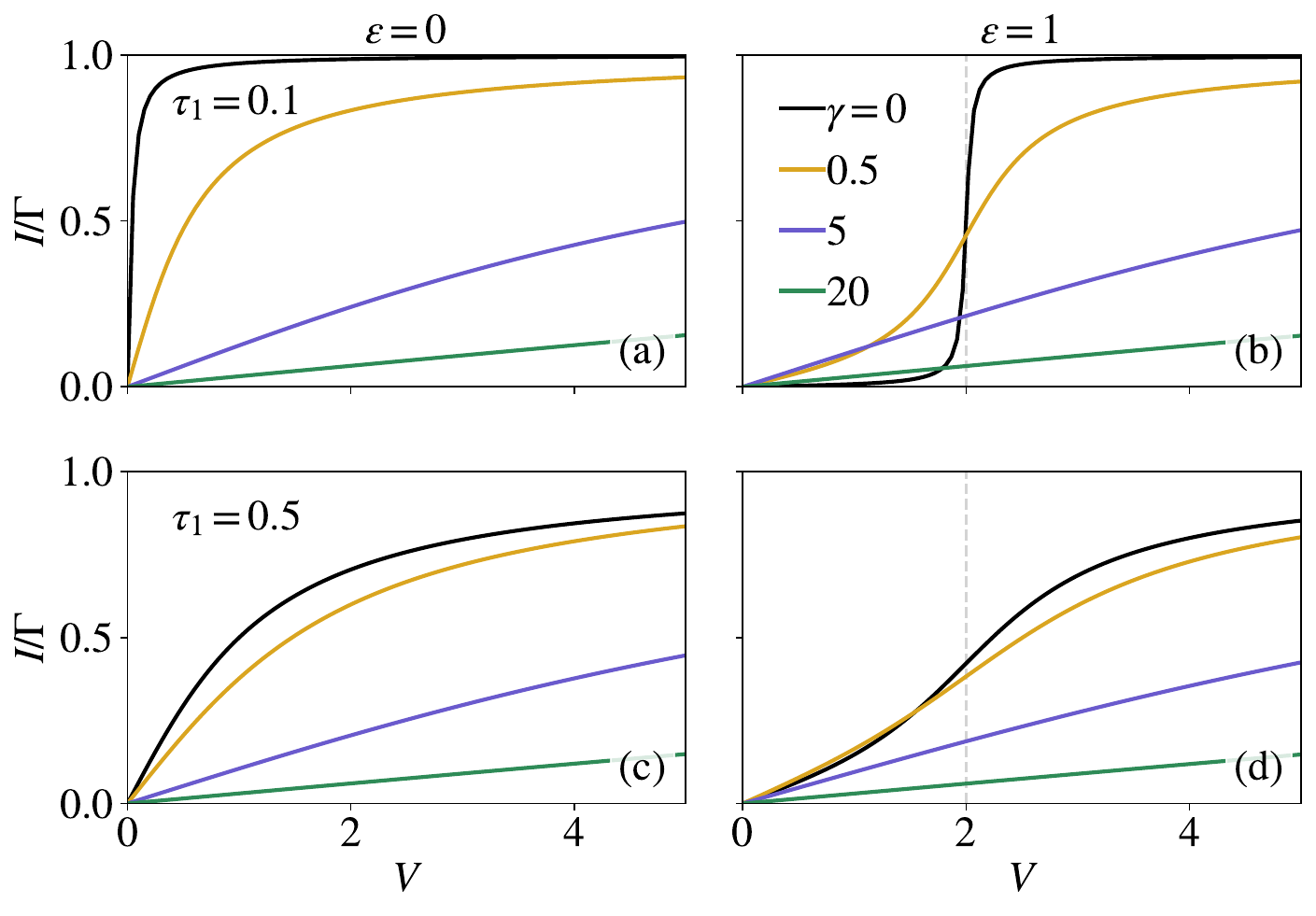}
\caption{Conserved current $I$ through a quantum dot coupled to leads as a function of voltage for (a, c) $\epsilon = 0$ and (b, d) $\epsilon = 1$. The current is scaled by the coupling $\Gamma = \pi \rho_0 \tau_1^2$. The tunneling amplitude is $\tau_1 = 0.1$ in panels (a, b) and $\tau_1 = 0.5$ in panels (c, d). The vertical line in in panels (b) and (d) marks voltage at which the chemical potential of the left reservoir coincides with the quantum dot energy level. Here, $V$, $\gamma$, and $\epsilon$ are in units of $\tau$, $\tau_1$ is in units of $\tau \sqrt{V}$, and $\tau = 1/(\pi \rho_0 \mathcal{V})$.}
\label{fig:conserved_current_qd}
\end{figure}

Using Eqs.~(\ref{eq:conserved_current_integral}) and~(\ref{eq:quantum_dot_infinite_cutoff}), the conserved current at zero temperature is
\begin{equation}
I = \frac{\Gamma}{\pi} \left[ \arctan\left( \frac{V - 2 \epsilon}{\gamma + 4\Gamma} \right) + \arctan\left( \frac{V + 2 \epsilon}{\gamma + 4\Gamma} \right) \right].
\end{equation}
Figure~\ref{fig:conserved_current_qd} shows the conserved current as a function of the applied voltage for a quantum dot. In the symmetric case $\epsilon=0$, the chemical potential of the quantum dot lies in the middle of the chemical potentials of the leads. In the absence of dissipation, the current increases quickly with voltage and then saturates to $I = \Gamma$. The broadening due to coupling to the reservoirs $\tau_1$ leads to a slower increase of the current with voltage, as seen in Fig.~\ref{fig:conserved_current_qd}(c).
The influence of the losses on the current is drastic. With increasing amplitude of the losses on the quantum dot, the current is strongly reduced for the voltages shown here. This reduction of the current at low voltages stems from the effective broadening of the energy level of the quantum dot by the dissipation. At large values of $\gamma$, there is only a slow, almost linear rise. However, at infinite voltage, the current saturates to $I = \Gamma$ independently of the finite loss rate $\gamma$.

The current-voltage curves change significantly in the case of an energy offset $\epsilon = 1$, shown in panels~(b) and~(d). In this situation, the curves have a step at $V = 2 \epsilon$, where the chemical potential of the left reservoir coincides with the chemical potential of the quantum dot. Both the broadening induced by the dissipation and the coupling to the reservoirs smoothen out the step-like feature and lead to a slow rise of the current.

\subsection{Three or more sites}

\begin{figure}[h!]
\includegraphics[width=\linewidth]{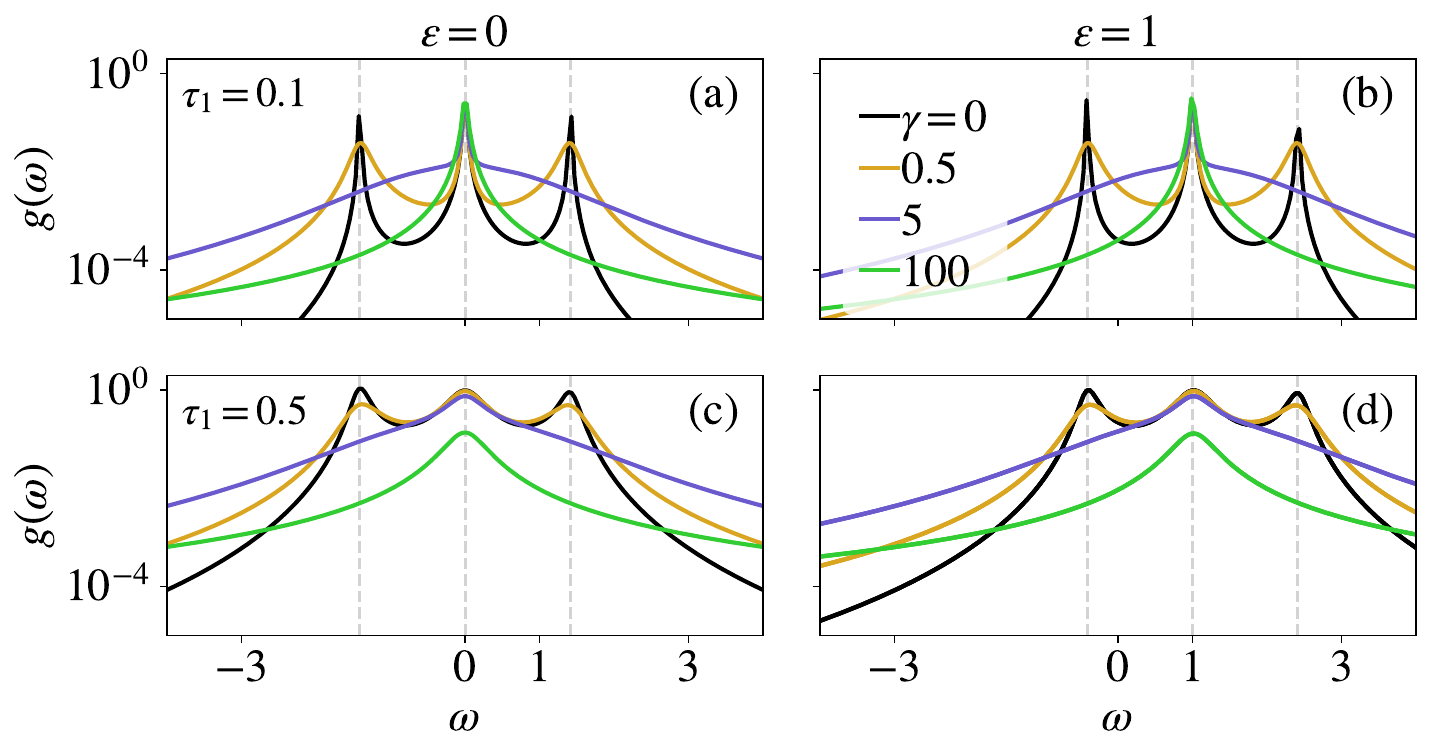}
\caption{Integrand $g(\omega)$ of Eq.~(\ref{eq:conserved_current_integral}) for the three-site system. There are three maxima approximately at the eigenenergies of the isolated system, marked by dashed vertical lines. Here, $\epsilon = 0$ in the left column and $\epsilon = 1$ in the right, $\tau_1 = 0.1$ in panels (a, b) and $\tau_1 = 0.5$ in panels (c, d).}
\label{fig:integrand_three_sites}
\end{figure}

For a lattice with $M$ sites, the existence of $M$ eigenstates leads to more complex current-voltage characteristics. This is connected to the structure of the integrand $g(\omega)$, which we plot for a three-site system in Fig.~\ref{fig:integrand_three_sites}. The function has three maxima at frequencies $\omega = E_0$ and $\omega \approx E_{\pm}$, where $E_0 = \epsilon$ and $E_{\pm} = \epsilon \pm \sqrt{2}\tau$ are the eigenenergies of an isolated three-site system not coupled to leads. The positions of the outer peaks are shifted by the coupling to the reservoirs: For $\gamma = 0$, these peaks are shifted to $\omega = \epsilon \pm \sqrt{2 \tau^2 - \Gamma^2}$ and are simultaneously broadened.

A nonzero dissipation rate $\gamma$ leads to a further shift and broadening, as shown in Fig.~\ref{fig:integrand_three_sites}.
However, a different broadening arises for the different peaks: The outermost peaks at $\omega \approx E_{\pm}$ are reduced and broadened much more than the central one at $\omega = \epsilon$. This is related to the symmetry of the isolated eigenstates pointed out in Ref.~\cite{VisuriKollath2022} in the context of the conductance. The eigenstates of an isolated lattice are either symmetric or antisymmetric, having either a finite overlap or a node at the center site, respectively. As the dissipation takes place at the center site, particles occupying the isolated antisymmetric eigenstates are not depleted by the loss. For the three-site system, the eigenstate with energy $\epsilon$ is antisymmetric, and therefore the corresponding central peak is reduced much less by the dissipation than the outermost peaks which arise from the symmetric eigenstates.

\begin{figure}[h!]
\includegraphics[width=\linewidth]{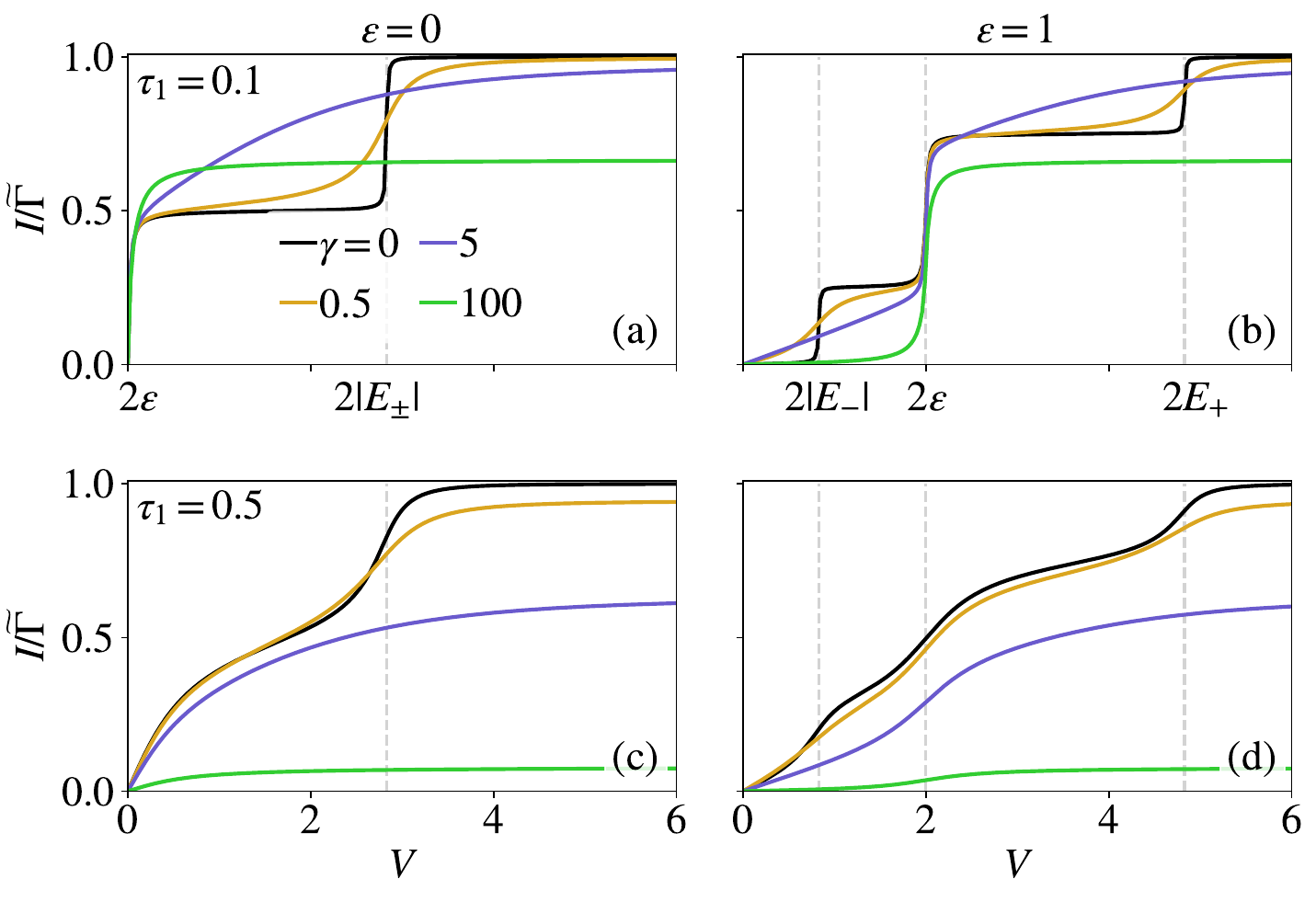}
\caption{Conserved current $I$ as a function of voltage in a three-site system for (a, c) $\epsilon = 0$ and (b, d) $\epsilon = 1$. The current is scaled by $\widetilde{\Gamma} = \Gamma \tau^2/(\Gamma^2 + \tau^2)$. In panels (a, b), $\tau_1 = 0.1$, and in (c, d), $\tau_1 = 0.5$. The units of the parameters are the same as in Fig.~\ref{fig:conserved_current_qd}. The current has a step-like structure with steps approximately at voltages $2 |E_n|$, marked by dashed vertical lines, where $E_n$ are the single-particle eigenenergies of an isolated system. For large dissipation $\gamma$, the steps arising from symmetric eigenstates disappear, while a larger coupling $\tau_1$ leads to a smoothening of all steps.}
\label{fig:conserved_current_3sites}
\end{figure}

\begin{figure}[h!]
\includegraphics[width=\linewidth]{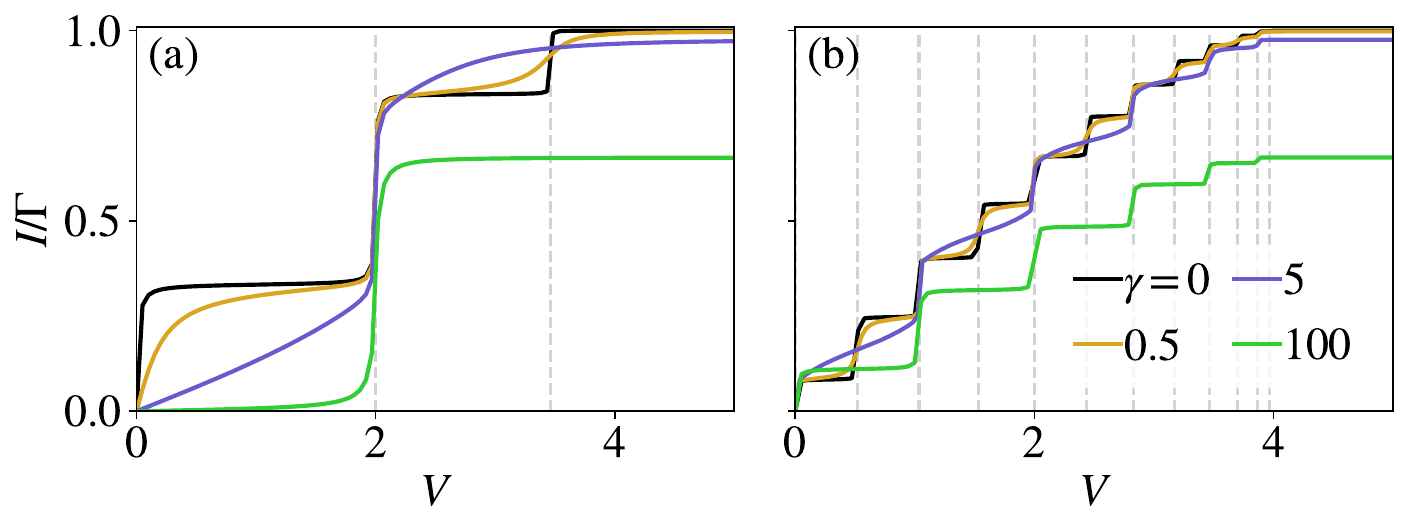}
\caption{Conserved current $I$ as a function of voltage in a system of (a) five (b) 23 sites, with $\epsilon = 0$ and $\tau_1 = 0.1$. The saturation current at $\gamma = 0$ is close to $\Gamma$ and is scaled by this value.}
\label{fig:conserved_current_5sites}
\end{figure}

The resonance structure due to single-particle eigenstates leads to a characteristic voltage dependence of the conserved current. In the absence of dissipation, the current-voltage curve has multiple pronounced steps at voltages corresponding approximately to the single-particle eigenenergies of the isolated system, as seen in Fig.~\ref{fig:conserved_current_3sites} (a, b) and Fig.~\ref{fig:conserved_current_5sites}. More precisely, a step occurs approximately when the chemical potential in either the left or the right lead coincides with one of the eigenenergies in the isolated chain. These positions of the discontinuities are exact only in the $\tau_1 \to 0$ limit, and a larger coupling to the leads induces a broadening and shift of the steps, seen in~Fig.~\ref{fig:conserved_current_3sites} (c, d). The conserved current saturates at large voltages. For the three-site system, in the absence of dissipation, the saturation value is obtained as $\widetilde{\Gamma} = \Gamma \tau^2/(\Gamma^2 + \tau^2)$.

The current-voltage curves are dramatically altered by the dissipation. 
Different effects occur depending on which eigenstates contribute to the transport, arising from the modification of the integrand $g(\omega)$. This can be seen in Fig. \ref{fig:conserved_current_3sites} where some of the steps are broadened and reduced more than others. Concentrating on $\epsilon=0$, we can compare the step arising at $V = 0$ for odd and even $l$. The odd-$l$ case is seen e.g. in Figs.~\ref{fig:conserved_current_3sites}(a) and~\ref{fig:conserved_current_5sites}(b) for three and 23 sites, and the even-$l$ case in Figs.~\ref{fig:conserved_current_qd}(a) and~\ref{fig:conserved_current_5sites}(a) for the quantum dot and five sites. Whereas for odd $l$, the step is relatively robust even if a large dissipation is applied, the corresponding step for even $l$ is smoothened out and nearly disappears with large dissipation. A similar effect is observed for the steps at higher voltages, where every second step is preserved while the others are suppressed by the dissipation. This behavior is connected to the form of the integrand $g(\omega)$, where the peaks corresponding to symmetric eigenstates are reduced by the dissipation much more than the ones corresponding to antisymmetric eigenstates. The broadening of the peaks also leads to a nonmonotonic dependence of the current on $\gamma$ at certain voltages. 

\begin{figure}[h!]
\includegraphics[width=0.8\linewidth]{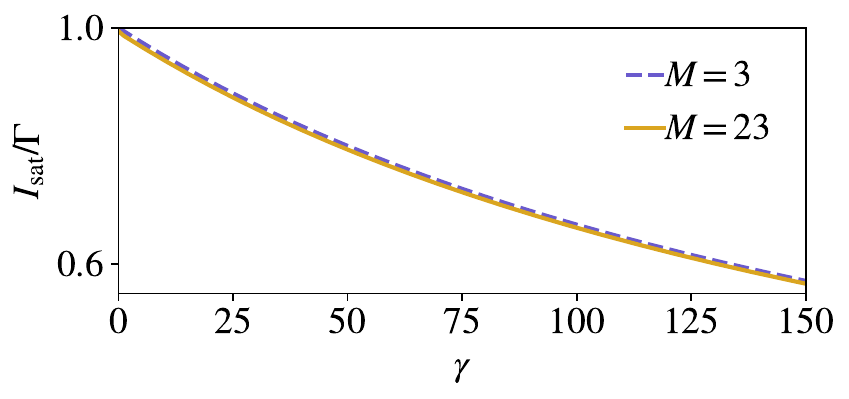}
\caption{For three or more sites, the saturation value of the current at large voltage decays with $\gamma$. For $M = 23$, the value of the current at $V = 5$ is used as $I_{\text{sat}}$. We set here $\tau_1 = 0.1$.}
\label{fig:saturation_decay}
\end{figure}

The saturation value of the current $I_{\text{sat}}$ at large voltage decreases in the presence of dissipation, decaying as $\sim 1/\gamma$ for $\gamma \gg \Gamma$. Figure~\ref{fig:saturation_decay} shows the saturation current as a function of dissipation rate, which is found to be approximately the same for lattices of three or more sites. For three sites, $I_{\text{sat}}$ is obtained analytically as the limit $\lim_{V \to \infty} I$, while for larger lattice sizes, we use the value of the current at a voltage which is larger than the lattice bandwidth. The maximum of $I_{\text{sat}}$ at zero dissipation is very close to $\Gamma$ when $\Gamma \ll \tau$.

\section{Loss current}

In this section, we first discuss the dependence of the loss current on the dissipation rate $\gamma$. In the first subsection, we focus on the symmetric case $\epsilon = 0$, where the loss current is independent of voltage, as explained in Appendix~\ref{app:loss_current}. The loss current displays a counterintuitive nonmonotonic dependence on the dissipation rate, known as the quantum Zeno effect.
Secondly, we focus on the voltage dependence of the loss current in the case of a nonzero energy offset $\epsilon$. 

The loss current, defined in Eq.~(\ref{eq:loss_current}), is proportional to the occupation of the lossy site and thus is given by the integral
\begin{equation}
I_{\text{loss}} = \gamma \int_{-\infty}^{\infty} \frac{d \omega}{2 \pi} f(\omega) \left[ n_L(\omega) + n_R(\omega) \right].
\label{eq:occupation_integral}
\end{equation}
Since the conserved current $I$ of Eq.~(\ref{eq:conserved_current_integral}) depends on the difference of the two Fermi distributions, and the chemical potential in either reservoir limits the range of the integration, $I$ has no strong dependence on the value of the cutoff when $V < \Lambda$. The loss current instead depends on the sum of the two Fermi distributions, and therefore the occupation and loss current potentially have a stronger dependence on the cutoff. In this section, we therefore analyze in detail the effect of a finite cutoff.

\subsection{Nonmonotonic dependence on $\gamma$}

A chain of atoms with a local particle loss has been shown to display the so-called quantum Zeno effect~\cite{BarmettlerKollath2011,BarontiniOtt2013,FromlDiehl2019,WolffKollath2020,FromlDiehl2020,MullerDiehl2021}, where the loss current behaves non-monotonically with the dissipation strength. Whereas for weak dissipation, the loss current is proportional to the loss amplitude, at large dissipation the loss current paradoxically becomes inversely proportional to $\gamma$. This is because in the $\gamma\to\infty$ limit, the tunneling to the dissipative site is strongly suppressed due to the energy mismatch between the neighboring sites.

\begin{figure}[h!]
\includegraphics[width=\linewidth]{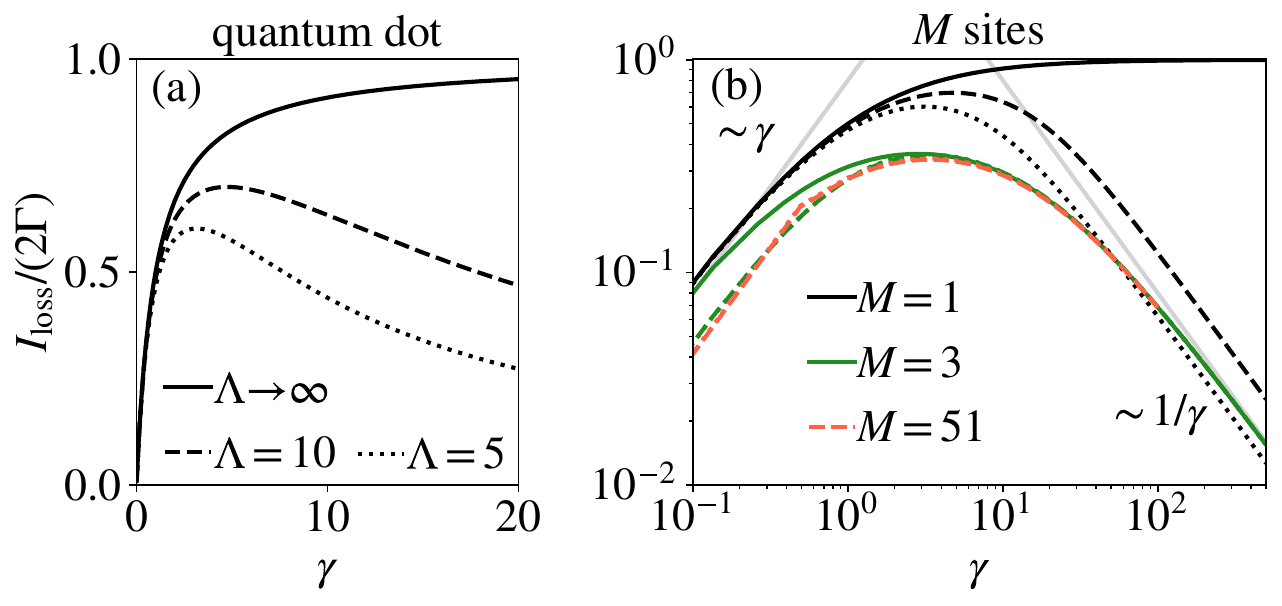}
\caption{The loss current as a function of $\gamma$ at $V = \epsilon = 0$ for different system sizes and cutoffs $\Lambda$. (a) For the quantum dot, the $\Lambda \to \infty$ limit is given by Eq.~(\ref{eq:saturation}). In this limit, the loss current saturates at $2 \Gamma$, while for finite $\Lambda$, there is a nonmonotonic dependence. (b) For three or more sites, $I_{\text{loss}}$ is nonmonotonic independently of the value of the cutoff. The solid, dashed, and dotted lines mark the different values of $\Lambda$ as in panel (a). The many-site system and the quantum dot with a finite cutoff have the same $\sim 1/\gamma$ dependence at large~$\gamma$, where the curves for different lattice sizes and different values of $\Lambda$ overlap. We set here $\tau_1 = 0.5$.} 
\label{fig:loss_current}
\end{figure}

The origin of the quantum Zeno effect as an energy mismatch can be exemplified by the system of a single quantum dot coupled to reservoirs.
In this case, the function $f(\omega)$ in Eq.~(\ref{eq:occupation_integral}) has the form
\begin{equation}
f(\omega) = \frac{8 \Gamma}{\left( \gamma + 4 \Gamma \right)^2 + 4(\omega - \epsilon)^2}
\label{eq:loss_current_integral}
\end{equation}
in the $\Lambda \to \infty$ limit, and the integral~(\ref{eq:occupation_integral}) evaluates to 
\begin{align}
\begin{split}
I_{\text{loss}} = \frac{2\Gamma \gamma}{\pi \left( \gamma + 4\Gamma \right)} \Big[\pi + &\arctan\left( \frac{V - 2 \epsilon}{\gamma + 4\Gamma} \right) \\
&- \arctan\left( \frac{V + 2 \epsilon}{\gamma + 4\Gamma} \right) \Big].
\end{split}
\label{eq:loss_current_voltage}
\end{align}
For $V = 0$ and $\epsilon = 0$, this simplifies to
\begin{equation}
I_{\text{loss}} = \frac{2 \Gamma \gamma}{\gamma + 4 \Gamma}.
\label{eq:saturation}
\end{equation}
For an infinite cutoff $\Lambda$, the loss current therefore saturates at large $\gamma$ with the saturation value $2 \Gamma$. Thus, when the energy of the particles tunneling to the quantum dot is unbounded, there is no energy mismatch which would suppress tunneling even for $\gamma \to \infty$, and the quantum Zeno effect does not occur. The nonmonotonic dependence is recovered if a finite cutoff is imposed. 
This is shown in Fig.~\ref{fig:loss_current}(a) where the loss current is plotted for different values of the cutoff. The value of $\gamma$ at which it has its maximum depends on the cutoff, and the quantum Zeno effect is only present if this energy scale is smaller than the dissipative coupling. 

For a lattice of three or more sites, we find a nonmonotonic dependence for all values of the cutoff, even in the $\Lambda \to \infty$ limit, as shown in Fig.~\ref{fig:loss_current}(b). The additional lattice sites in this case provide an effective cutoff on the energies from which particles can tunnel to the lossy site. The loss current can be obtained analytically in the $\Lambda \to \infty$ limit when the lattice size is small, and we plot the curves with both finite and infinite cutoff for three sites. For the largest lattice size $M = 51$ we only show the numerical solution in the case of finite $\Lambda$. For different values of $\Lambda$, the curves overlap for $\gamma \gtrsim 10$, and the same dependence $I_{\text{loss}} \propto \gamma^{-1}$ is recovered for different lattice sizes $M \geq 3$ as for the quantum dot system with a finite cutoff.

\subsection{Voltage dependence}
\label{sec:loss_current-voltage}
\begin{figure}[h!]
\includegraphics[width=\linewidth]{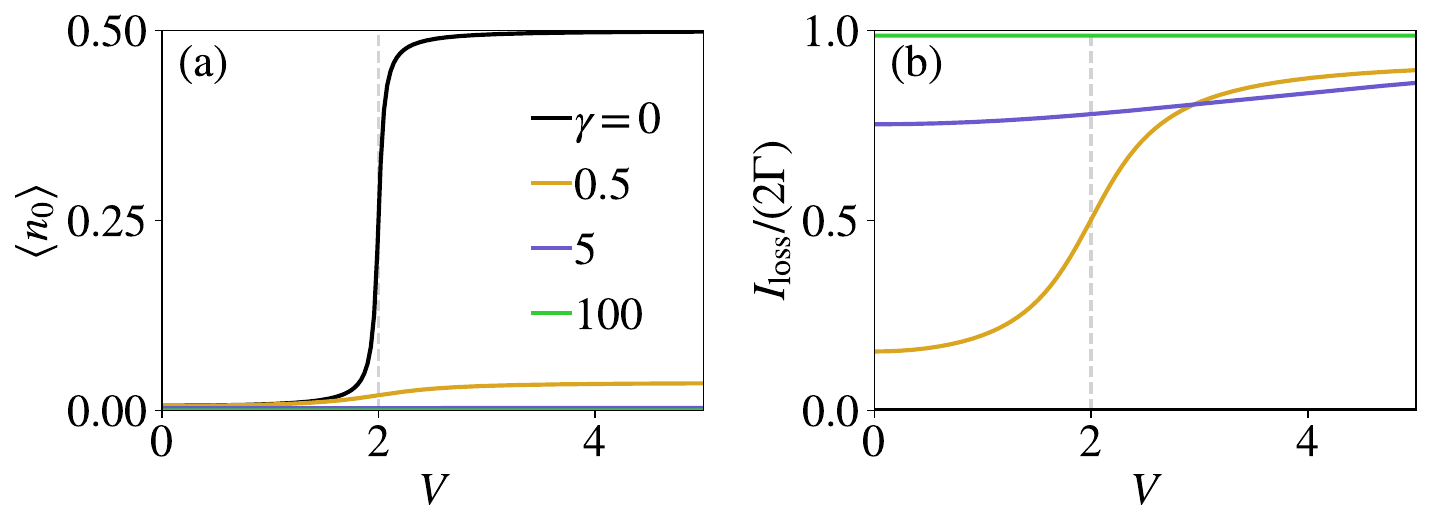}
\includegraphics[width=\linewidth]{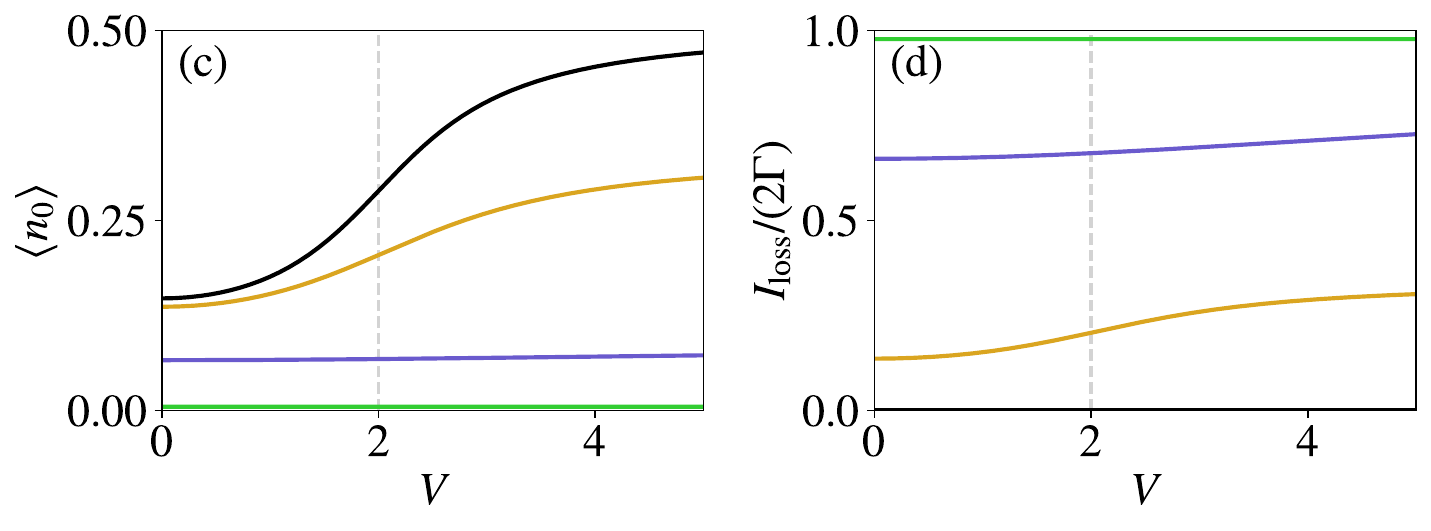}
\caption{The quantum dot occupation (a, c) and the loss current (b, d) as functions of voltage for different values of $\gamma$. In panel~(a), the lines for $\gamma = 5$ and $\gamma = 100$ are very close to zero, and in panels~(b) and~(d), the loss current is zero for $\gamma = 0$. Here, $\epsilon = 1$ and the coupling to the reservoirs is $\tau_1 = 0.1$ (a, b) and $\tau_1 = 0.5$ (c, d). The vertical lines mark the voltage at which $\mu_L$ coincides with $\epsilon$.}
\label{fig:loss_current_qd}
\end{figure}

We find that the loss current for an energy offset $\epsilon = 0$ is independent of the voltage (see Appendix~\ref{app:loss_current_voltage}). However, in the presence of an energy offset $\epsilon \neq 0$, the loss current acquires a dependence on the voltage. Figure~\ref{fig:loss_current_qd} shows the occupation of a quantum dot coupled to reservoirs and the loss current from the dot. For $\epsilon = 1$, we see that the quantum dot is nearly empty when the chemical potential of the left reservoir is below~$\epsilon$.
A step-like increase in the occupation occurs when $\mu_L = V/2$ coincides with~$\epsilon$. The quantum dot becomes occupied and at larger voltages, the occupation approaches one-half. This change is reflected in the loss current $I_{\text{loss}}$, which shows a step-like dependence on voltage for intermediate dissipation rates.
Mathematically, the existence and position of the step can be understood from Eq.~(\ref{eq:loss_current_integral}), where similarly to $g(\omega)$ in Eq.~(\ref{eq:conserved_current_integral}), the function $f(\omega)$ has a maximum at~$\epsilon$. 

The step-like change is the most pronounced for a small coupling $\tau_1$ [Fig.~\ref{fig:loss_current_qd}(a, b)].
When the coupling to the leads is larger, the occupation of the dot changes more smoothly since particles can more easily tunnel in and out of the dot. 
This results in a wider step in the loss current as seen in Fig.~\ref{fig:loss_current_qd}(c,~d). 
For large $\gamma$, one can see that the integrand $f(\omega)$ is broadened into a constant and the step is smoothened out completely.
We consider here the limit $\Lambda \to \infty$, where the loss current saturates at large $\gamma$. The case of a finite cutoff is presented in Appendix~\ref{app:finite_continuum} for completeness. 

\begin{figure}[h!]
\includegraphics[width=\linewidth]{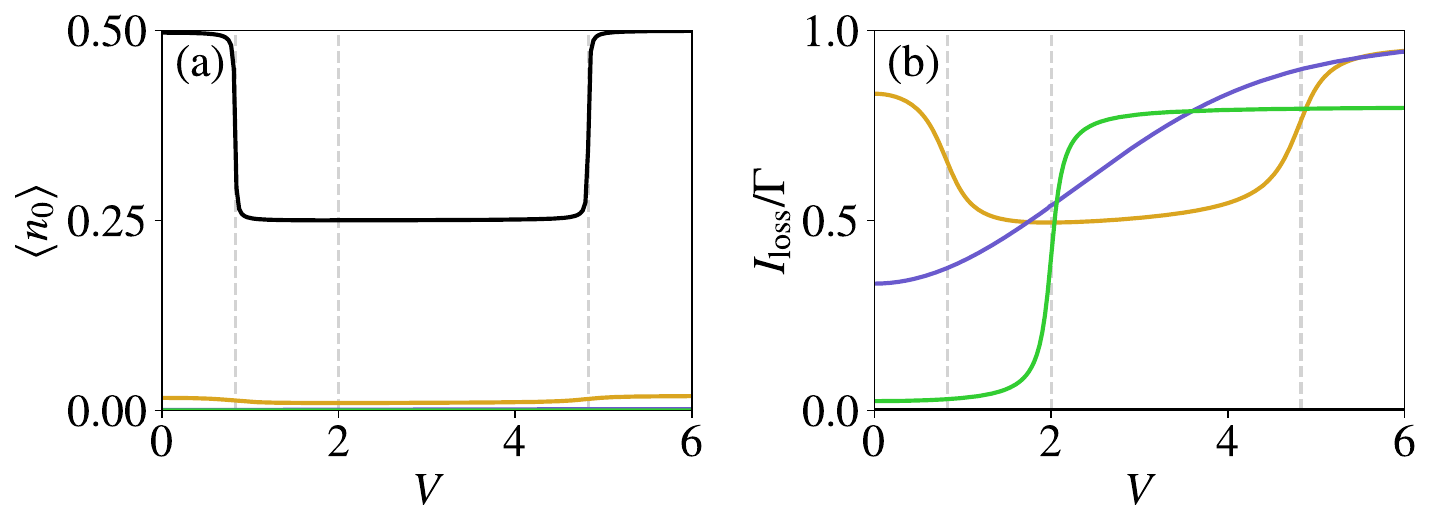}
\includegraphics[width=\linewidth]{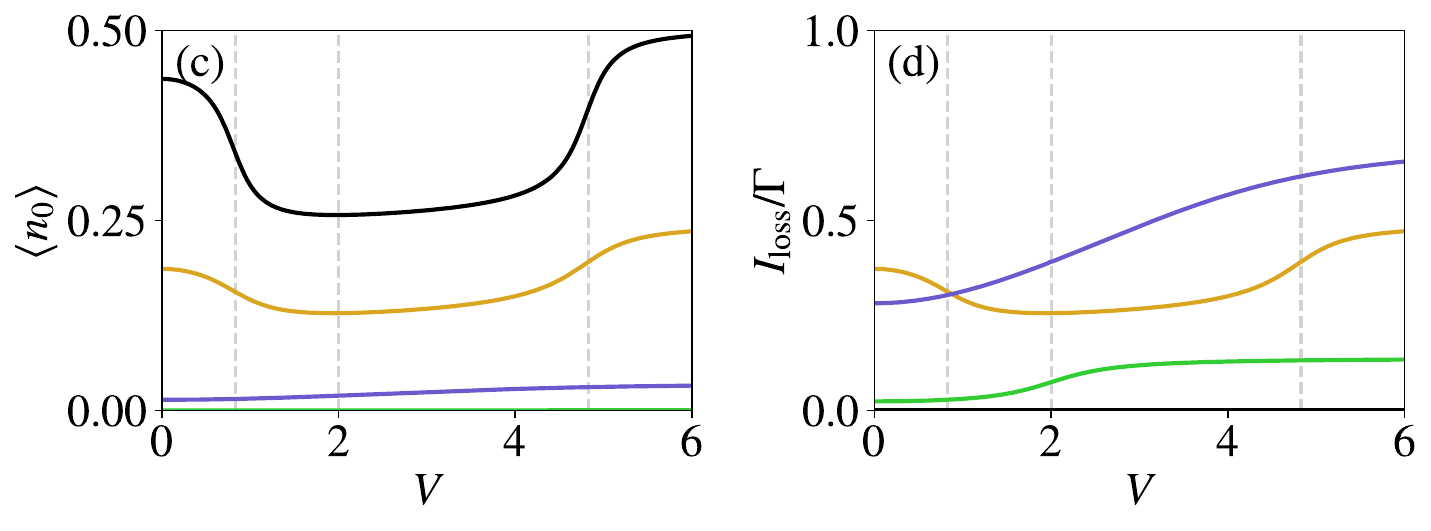}
\caption{The occupation of the lossy site and the loss current in a three-site system with $\epsilon = 1$, (a, b) $\tau_1 = 0.1$ and (c, d) $\tau_1 = 0.5$. The line colors are as in Fig.~\ref{fig:loss_current_qd}. The vertical lines mark the voltages $V/2 = |E_n|$, where $E_n$ are the single-particle eigenstates in the $\tau_1 \to 0$ limit.}
\label{fig:loss_current_three_sites}
\end{figure}

For three sites and an energy offset $\epsilon = 1$, a more complex voltage dependence appears due to the resonances of the reservoir chemical potential with single-particle eigenenergies. This is shown in Fig.~\ref{fig:loss_current_three_sites}. 
At zero voltage, without dissipation, the center site is half-filled as the chemical potential in the reservoirs is above the lowest eigenenergy of the lattice. The corresponding eigenstate is one of the two symmetric eigenstates which have a nonzero overlap with the lossy site and contribute to the occupation on that site. With increasing voltage, the chemical potential of the right reservoir becomes lower than the lowest eigenstate energy, leading to a drop in the center site occupation to approximately $0.25$ at $V = 2|\epsilon - \sqrt{2} \tau|$. When the voltage further increases, the chemical potential on the left exceeds the highest eigenenergy corresponding to the other symmetric eigenstate, and the occupation increases to one-half again. 
For $\tau_1 = 0.1$, shown in Fig.~\ref{fig:loss_current_three_sites}(a), there is no visible change in the occupation at $V = 2 \epsilon$ where $\mu_L$ crosses the eigenenergy in the middle of the spectrum. This is because the corresponding eigenstate is antisymmetric and does not contribute to the occupation of the center site in the $\tau_1 \to 0$ limit. For larger $\tau_1$, as in Fig.~\ref{fig:loss_current_three_sites}(b), the eigenstates are modified so that the centermost eigenstate develops a finite overlap with the lossy site. This leads to a small increase of the occupation $\braket{n_0}$ between voltages $V = 2|\epsilon - \sqrt{2} \tau|$ and $2(\epsilon + \sqrt{2}\tau)$.

When the particle losses act, the center site occupation is significantly reduced already for small dissipation, making the distinctive features almost invisible. One can see that for a weak coupling to the leads, the dissipation is much more effective in depleting the central site [Fig.~\ref{fig:loss_current_three_sites}(a)] than at larger coupling [Fig.~\ref{fig:loss_current_three_sites} (b)]. In the loss current, an interesting change in behavior can be observed when the dissipation rate increases. 
For small dissipation $\gamma = 0.5$, the loss current has the same decrease and subsequent increase as the center site occupation in the absence of dissipation, determined only by the symmetric eigenstates. However, this dependence changes crucially for larger values of $\gamma$, where the antisymmetric eigenstate also becomes of importance. The voltage dependence changes into a single step centered around $V = 2 \epsilon$, corresponding to the eigenenegy of the antisymmetric eigenstate, as seen in Figs.~\ref{fig:loss_current_three_sites}(b, d).
This is due to the fact that for large $\gamma$, the occupation of the symmetric eigenstates is almost completely depleted, while particles in the antisymmetric eigenstate are less affected. Therefore, features arising from the antisymmetric eigenstate become visible, causing this drastic change in the loss current as a function of dissipation.

\section{Momentum distribution and Friedel oscillations}
\label{sec:friedel_oscillations}

For larger systems, it is interesting to study not only the occupation of the center site but also of the remaining lattice. While diffusive transport, such as in metallic wires~\cite{PothierDevoret1997}, leads to a linear change in the steady-state particle density, and the transport of free fermions is ballistic with a uniform density distribution, the local particle loss creates a density drop across the lossy site~\cite{VisuriKollath2022}. 
The latter situation is different from what is typical for either diffusive or ballistic transport.
In addition to the density drop, the density distribution shows interesting features 
connected to the momentum distribution, and we present both in this section. In the first subsection, we concentrate on $\epsilon = 1$ where the average filling is one-third in the lossless lattice, and in the second subsection, we discuss the half-filled lattice with $\epsilon = 0$.

\subsection{One-third filling}

\begin{figure}[h!]
\includegraphics[width=\linewidth]{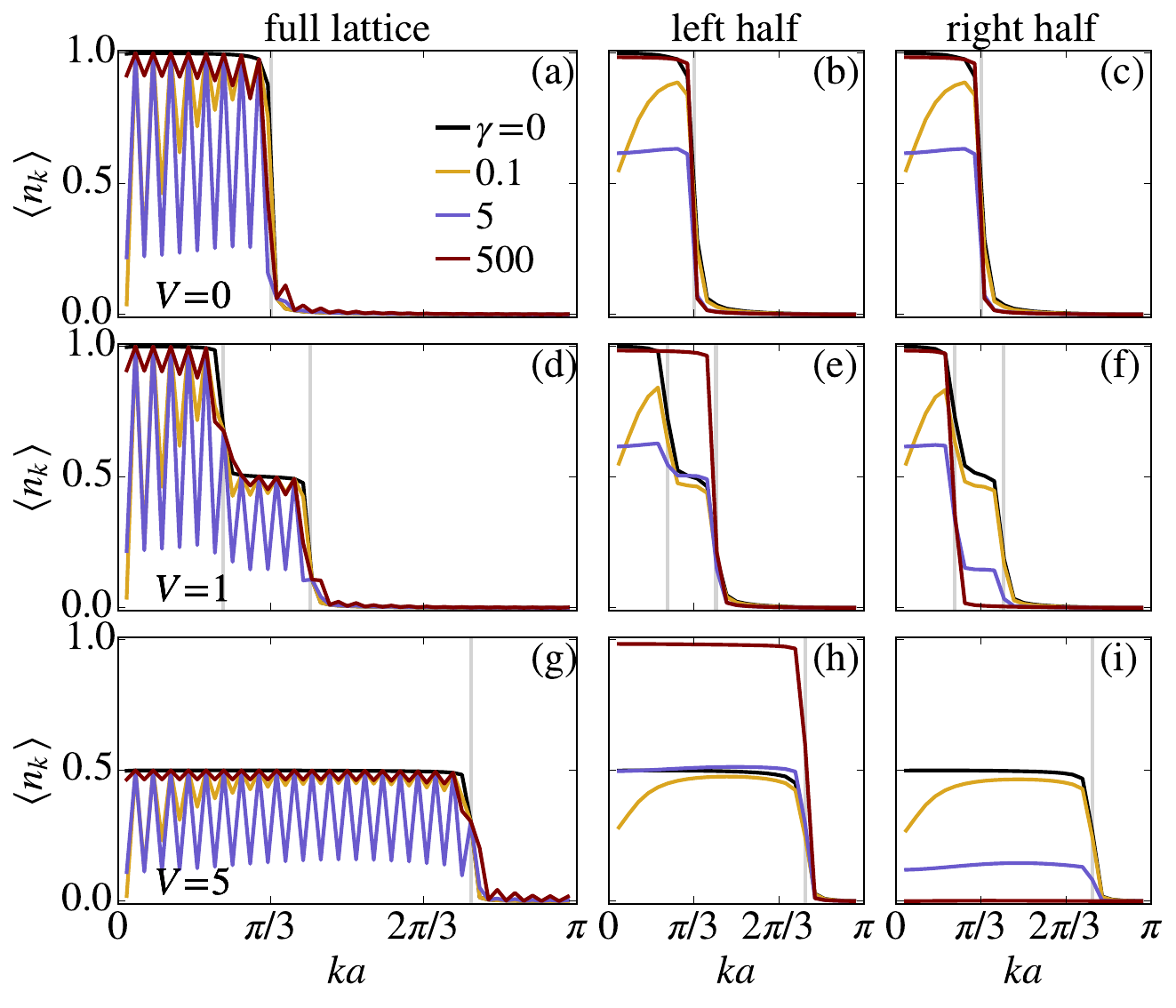}
\caption{The quasimomentum distribution $\braket{n_k}$ in the lattice at different dissipation rates. The leftmost column [panels (a, d, g)] shows $\braket{n_k}$ in the full lattice, as given by Eq.~(\ref{eq:momentum_distribution}). The middle and rightmost columns correspond to the left and right halves of the lattice, with momentum distribution given by Eqs.~(\ref{eq:left_momentum}) and~(\ref{eq:right_momentum}), respectively. On the first row (a-c), the voltage is $V = 0$, on the second row (d-f), $V = 1$, and on the third row (g-i), $V = 5$.  The Fermi momenta given by Eq.~(\ref{eq:fermi_momentum}) are marked by vertical gray lines. The lattice chemical potential is set to $\epsilon = 1$, so that at zero voltage and in the absence of dissipation, the lattice filling is $1/3$.}
\label{fig:momentum_one-third_filling}
\end{figure}

The momentum distribution is drawn in Fig.~\ref{fig:momentum_one-third_filling}. The first column of the figure shows the distribution in the full lattice, given by Eq.~(\ref{eq:momentum_distribution}), and the second and third columns correspond to the left and right halves excluding the lossy site. The momentum distribution in the left ($L$) and right ($R$) halves is given by
\begin{align}
\braket{n_{k L}} &= \sum_{i, j = 1}^{(M-1)/2} \varphi_{i, k} \varphi_{j, k} \braket{d_i^{\dagger} d_j^{\phantom{\dagger}}}, 
\label{eq:left_momentum} \\
\braket{n_{k R}} &= \sum_{i, j = 1}^{(M-1)/2} \varphi_{i, k} \varphi_{j, k} \braket{d_{(M+1)/2+i}^{\dagger} d_{(M+1)/2+j}^{\phantom{\dagger}}},
\label{eq:right_momentum}
\end{align}
where $\varphi_{j, k} = (2/\sqrt{M+1}) \sin(k j)$ and the quasimomentum is discretized as $k = 2n\pi/(M+1)$ with $n \in \{1, 2, ..., (M-1)/2\}$.

When the chemical potentials of the reservoirs are equal, as on the first row, the lattice filling is one-third. The lowest momentum states are occupied up to the Fermi momentum $k_F \approx \pi/3$ in the absence of dissipation. For an isolated lattice at zero temperature, $\braket{n_k}$ would have a sharp discontinuity at $k_F$. Here, however, the momentum states are not the exact eigenstates due to the coupling to the reservoirs, and therefore the discontinuity is rounded. When $V = 0$, the momentum distributions are equal in the left and right halves of the lattice, as seen in Figs.~\ref{fig:momentum_one-third_filling}(b, c).
The main effect of the local dissipation is to deplete the occupation of the symmetric eigenstates which have a large overlap with the lossy site. This results in the minima at every second momentum value in panels~(a, d, g). For small loss rates ($\gamma = 0.1$), the depletion is strongest for the lowest-momentum eigenstate which has the largest amplitude at $j = 0$. The momentum distributions in the left and right halves are calculated in the basis of states where the wavefunction is zero at $j = 0$. They therefore exclude the states which are depleted by the dissipation and do not show an alternating pattern.
 
In the situation where the chemical potential is different in the left and right reservoirs but the voltage is smaller than the lattice bandwidth, such as $V~=~1$ in Fig.~\ref{fig:momentum_one-third_filling}(d-f), the momentum distribution of the dissipation-free system has two steps. Their positions coincide with the Fermi momenta that would exist in an equilibrium system where the lattice is coupled to only the left or right reservoir. These Fermi momenta in the lattice can be estimated by equating the chemical potentials in either reservoir with the Fermi energy $\varepsilon_F = \epsilon-2 \tau \cos(k_F)$ in the lattice. The Fermi momentum is then given by
\begin{equation}
k_{F, i} = \arccos \left(- \frac{\mu_i}{2 \tau} + \frac{\epsilon}{2 \tau}\right).
\label{eq:fermi_momentum}
\end{equation}

\begin{figure}[h!]
\includegraphics[width=\linewidth]{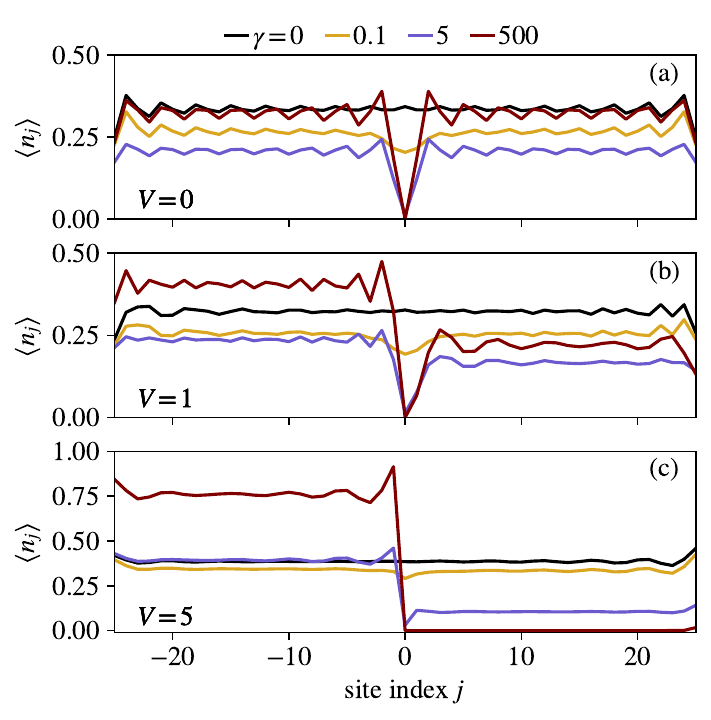}
\caption{Particle density $\braket{n_j}$ for a lattice of 51 sites with $\epsilon = 1$ and $\tau_1 = 0.5$ as in Fig.~\ref{fig:momentum_one-third_filling}. The average filling for $\gamma = 0$ is $1/3$. The density imbalance between the left and right sides develops as a combined effect of the finite voltage and dissipation.}
\label{fig:one-third_filling}
\end{figure}

A similar feature was measured in the energy distribution of quasiparticles in mesoscopic wires~\cite{PothierDevoret1997}, where two discontinuities appear at the Fermi levels of the leads. The height of the second discontinuity, however, changes across the wire whereas here it is fixed at $0.5$. This is because unlike in mesoscopic wires, where transport is diffusive and the electron density changes linearly across the wire, the free-fermion system studied here is ballistic in the absence of particle loss and the density is uniform. Figures~\ref{fig:momentum_one-third_filling}(e, f) also show that in the limit of strong dissipation, each half of the lattice develops a single Fermi momentum determined by the chemical potential of the reservoir on that side. This corresponds to an imbalance in the average density between the left and right halves~\cite{VisuriKollath2022}.
The particle density distribution in the lattice is plotted in Fig.~\ref{fig:one-third_filling}, where a nonzero voltage and a strong dissipation are indeed seen to give rise to a sharp density drop across the lossy site in panels~(b, c).

Apart from a minimum at the lossy site and an imbalance between the left and right sides at nonzero voltage and dissipation, we observe that the background density -- the particle density in the lattice away from the lossy site -- has a nonmonotonic dependence on the dissipation rate. This is seen most clearly in Fig.~\ref{fig:one-third_filling}(a): after an initial depletion of the density with $\gamma > 0$, the average density approaches the $\gamma = 0$ value in the limit of large~$\gamma$. A similar nonmonotonic behavior in the absence of a voltage was reported previously in Ref.~\cite{FromlDiehl2020} and has its origin in the quantum Zeno effect. When there is a chemical potential difference between the reservoirs, as in Figs.~\ref{fig:one-third_filling}(b, c), the limiting value of the average density does not approach the $\gamma = 0$ value anymore, but a density drop develops across the lossy site. In the $\gamma \to \infty$ limit, the density distribution is equal to that of two disconnected halves of the lattice, in each of which the filling is determined only by the density in the reservoir coupled to that half.

The presence of a boundary in a fermionic system typically leads to Friedel oscillations. 
In an equilibrium system, Friedel oscillations have a wavevector $2 k_F$, where $k_F = n_0 \pi$ is the Fermi momentum and $n_0$ the average density. This matches the wavevector seen in Fig.~\ref{fig:one-third_filling}(a) for $\gamma = 0$. Furthermore, we see that in the lossy system, the wavevector is approximately equal for different dissipation rates. The wavelength of the Friedel oscillations is therefore determined by the density of the reservoirs rather than the average density in the lattice. This is consistent with the observation that the local loss does not change the Fermi momentum in Fig.~\ref{fig:momentum_one-third_filling} but rather depletes alternating momentum states across the spectrum.

In the case $V = 1$, the wavevector of the Friedel oscillations is different in the left and right halves. For large $\gamma$, the wavevector on either side is $2 k_{F, i}$ with $k_{F, i}$ given by Eq.~(\ref{eq:fermi_momentum}): On the left side, the Fermi momentum is $k_{F, L} \approx 0.2 \pi$, so that the wavelegth of the oscillations is $\lambda_L \approx 2.4$. This matches the approximately $10$ wavelegths contained in $25$ sites in Fig.~\ref{fig:one-third_filling}(b). On the right, $k_{F, R} \approx 0.4 \pi$. The expected wavelength is $\lambda_R \approx 4.3$, which agrees with the $\sim 5$ wavelengths within 25 sites in the right half of the lattice.

Interestingly, in either half, Friedel oscillations occur with a different wavevector at $\gamma = 0$ than $\gamma = 500$, and it seems that the wavevectors are inverted between the left and right sides. This feature is not reflected by the left and right momentum distributions which at $\gamma = 0$ are nearly identical.
In Fig.~\ref{fig:one-third_filling}(c), the right half of the lattice is empty for $\gamma = 500$, while the left half is less than fully filled and displays Friedel oscillations with a wavevector given by the hole density. With the energy offset $\epsilon = 1$, the lattice filling in the $\gamma \to \infty$ limit reaches zero in the right half at $V = 2$, where the argument of the $\arccos$ function in Eq.~(\ref{eq:fermi_momentum}) is equal to one. A full filling of the left half is correspondingly reached at $V = 6$.

\subsection{Half filling}

\begin{figure}[h!]
\includegraphics[width=\linewidth]{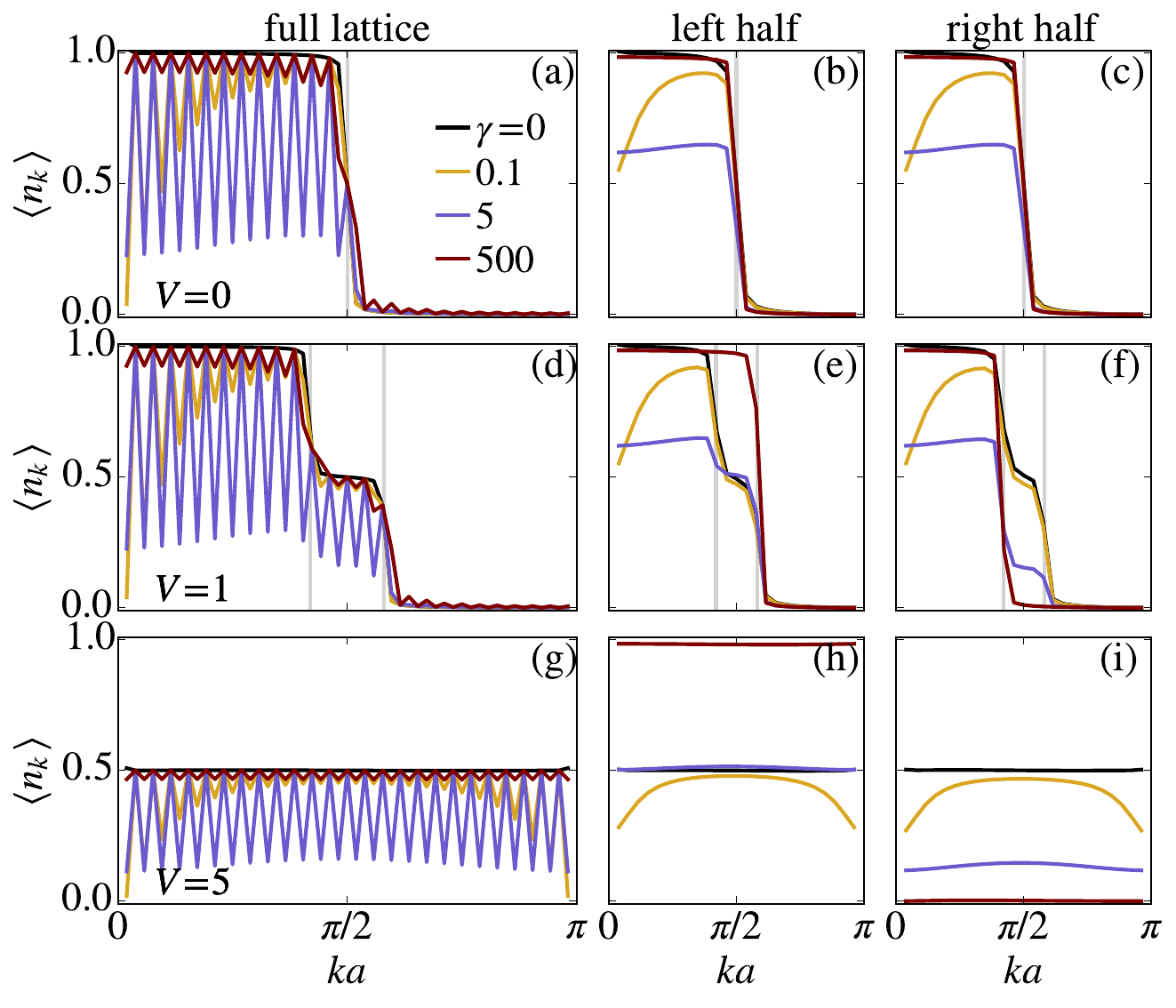}
\caption{The quasimomentum distribution $n_k$ in the lattice, as in Fig.~\ref{fig:momentum_one-third_filling}, for $\epsilon = 0$. The first row (a-c) corresponds to $V = 0$, the second row (d-f) to $V = 1$, and the third row (g-i) to $V = 5$. At zero voltage and in the absence of dissipation, the lattice is half-filled.}
\label{fig:momentum_half_filling}
\end{figure}

The momentum distribution for $\epsilon = 0$, where the lattice without losses is half-filled, is plotted in Fig.~\ref{fig:momentum_half_filling}. For equal chemical potentials in the reservoirs, the momentum states are filled up to $k_F = \pi/2$, while at $V = 1$, there are two discontinuities at the Fermi momenta given by Eq.~(\ref{eq:fermi_momentum}) with $\mu_{L, R} = \pm V/2$ and $\epsilon = 0$. 
When the voltage is larger than the bandwidth, such as in Figs.~\ref{fig:momentum_half_filling}(g-i), all momentum states are equally occupied in the absence of dissipation. For small dissipation $\gamma = 0.1$, the maximum depletion occurs for the lowest and highest momentum states symmetrically, while for stronger dissipation, the depletion of symmetric eigenstates is nearly uniform across the spectrum.

\begin{figure}[h!]
\includegraphics[width=\linewidth]{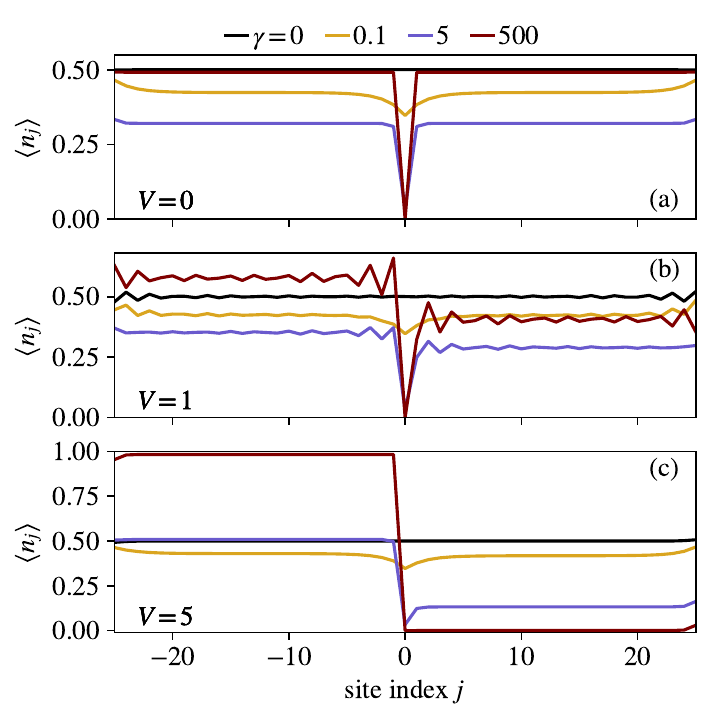}
\caption{Particle density $\braket{n_j}$ for a lattice of 51 sites as a function of the position $j$ for different voltages $V$ and losses $\gamma$ with $\epsilon = 0$ and $\tau_1 = 0.5$.}
\label{fig:half_filling}
\end{figure}

Figure~\ref{fig:half_filling}(a) shows the density distribution at $\epsilon = \mu_L = \mu_R = 0$, where the lossless lattice is half-filled and Friedel oscillations are suppressed due to particle-hole symmetry. Interestingly, Friedel oscillations are absent even for nonzero $\gamma$ when the steady-state particle density deviates from half filling. At a finite voltage $V = 1$, Friedel oscillations appear at the boundaries with the reservoirs and around the dissipative site. 
While in the right half of the lattice, the average density in the large-$\gamma$ limit is below one-half and the wavevector of the Friedel oscillations is determined by the particle density, on the left side, the average density is above one-half and the Friedel oscillations are governed by the hole density. They therefore have the wavevector $2 (1 - n_0) \pi = 2 k_{F, R}$. 
Figure~\ref{fig:half_filling}(c) shows the density distribution in the case where the voltage $V = 5$ is larger than the bandwidth $4 \tau$ with $\tau = 1$. For large dissipation $\gamma = 500$, left half of the lattice is nearly fully filled and the right half empty, apart from small deviations at the edges of the lattice. In the fully filled or empty system, there are no Friedel oscillations. We observe that the oscillations are also absent for smaller values of $\gamma$ where the average density of the lattice is close to $1/2$. 
This is connected to the absence of a Fermi momentum in Figs.~\ref{fig:momentum_half_filling}(g-i).

\section{Particle density imbalance}
\label{sec:imbalance}

\begin{figure}[h!]
\includegraphics[width=\linewidth]{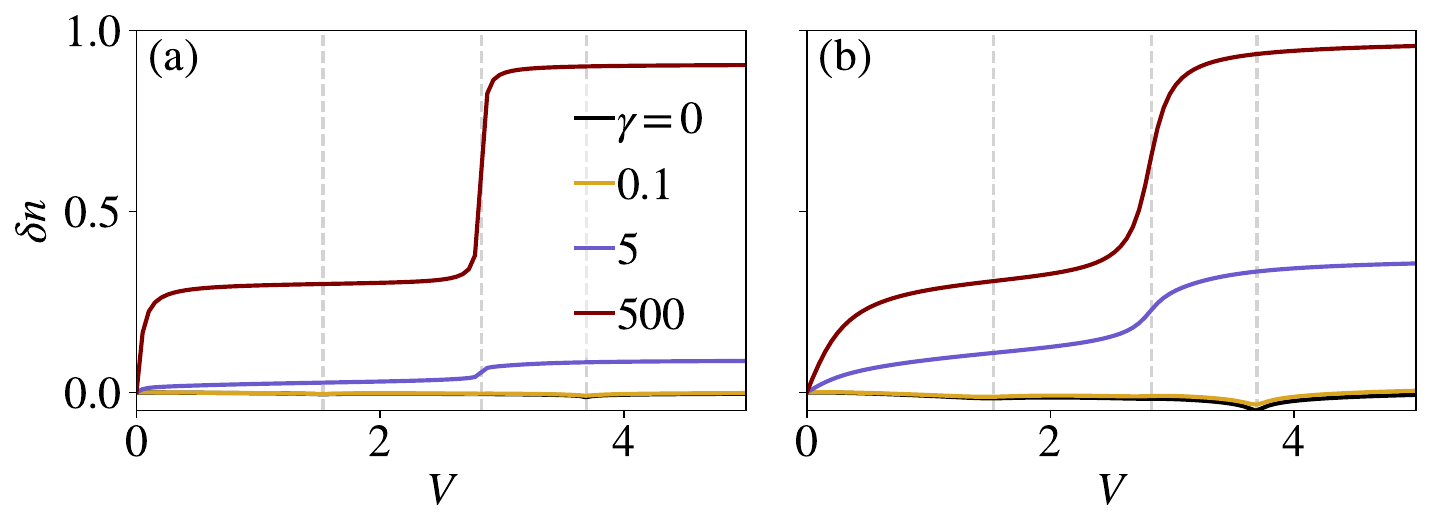}
\caption{The average imbalance $\delta n$ as a function of voltage in a lattice of seven sites, with (a) $\tau_1 = 0.1$ and (b) $\tau_1 = 0.5$. A larger coupling $\tau_1$ leads to the broadening of the steps. The vertical lines mark the values $V/2 = E_n$ where $E_n$ are the single-particle eigenstates in the $\tau_1 \to 0$ limit.}
\label{fig:imbalance_L7}
\end{figure}

In the presence of both a finite voltage and a particle loss, a density drop develops across the lossy site. This is seen in Figs.~\ref{fig:one-third_filling} and~\ref{fig:half_filling}. Here, we analyze the resulting average density imbalance $\delta n$ between the left and right halves of the lattice, excluding the lossy site,
\begin{equation}
\delta n = \braket{n_L} - \braket{n_R} = \frac{2}{M - 1} \left(\sum_{j<0} \braket{n_j} - \sum_{j>0} \braket{n_j} \right).
\label{eq:imbalance}
\end{equation}
We focus on the case $\epsilon = 0$, where the lattice is half-filled in the absence of loss. 
In Fig.~\ref{fig:imbalance_L7}, we plot the density imbalance as a function of voltage for the representative case of seven lattice sites. The imbalance has a step-like behavior similar to the conserved current and loss current discussed in Sections~\ref{sec:current-voltage} and~\ref{sec:loss_current-voltage}: For a sufficiently large loss, it grows in steps approximately when the chemical potential in either reservoir coincides with an eigenenergy of an isolated lattice. These steps however occur only for the energies of antisymmetric eigenstates, since symmetric eigenstates are depleted by the dissipation and do not contribute to changes in the average density. This can be seen in Fig.~\ref{fig:imbalance_L7}, where the steps occur at $\epsilon = 0$ and approximately at $\epsilon = \sqrt{2}\tau$. A larger coupling to the reservoirs leads, as for the currents, to a broadening of the steps. In Sec.~\ref{sec:equilibrium}, we show that the broadening can be reproduced by a simple model of a quantum dot coupled to a single reservoir at equilibrium.

\begin{figure}[h!]
\includegraphics[width=\linewidth]{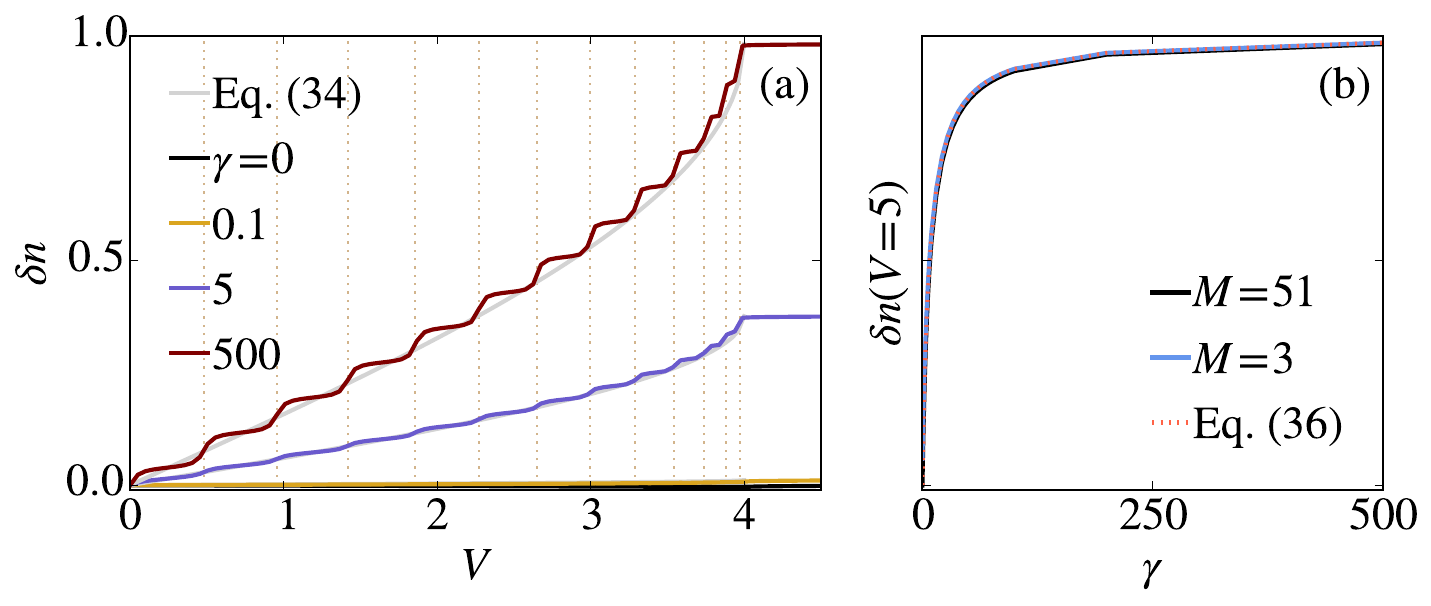}
\caption{(a) The average imbalance as a function of voltage for a 51-site lattice, calculated as in Eq.~(\ref{eq:imbalance}), for $\tau_1 = 0.5$. The voltages where $V/2$ coincides with the eigenvalues of antisymmetric eigenstates are marked with vertical lines. The overall slope matches the estimate $(k_{F, L} - k_{F, R})/\pi$ with $k_{F, i}$ given by Eq.~(\ref{eq:fermi_momentum}) (see text). (b) The saturation imbalance at $V = 5$ as a function of the dissipation rate coincides for 51 and 3 sites, and is reproduced by the simple formula~(\ref{eq:imbalance_expansion}).}
\label{fig:imbalance_L51}
\end{figure}

The imbalance is shown for a lattice of 51 sites in Fig.~\ref{fig:imbalance_L51}(a), where the antisymmetric eigenstate energies are marked by vertical lines. The imbalance saturates when the voltage exceeds the bandwidth of the lattice, with a saturation value that depends on $\gamma$. We find that the overall slope is very well reproduced by $(k_{F, L} - k_{F, R})/\pi$, with $k_{F, i}$ given by Eq.~(\ref{eq:fermi_momentum}), when this function is multiplied by the saturation value extracted from the numerical result. In panel~(b), we plot the saturation value at $V = 5$ as a function of the dissipation rate. The result for 51 sites coincides with the result for three sites. These results with fixed voltage are reproduced by the analytic limit $V \to \infty$ of the three-site lattice, which for large $\gamma$ can be approximated by (see Appendix~\ref{app:three_sites})
\begin{equation}
\lim_{V \to \infty} \delta n \approx \frac{\gamma}{\frac{2 \tau^2}{\Gamma} + \gamma}.
\label{eq:imbalance_expansion}
\end{equation}
A larger coupling to the reservoirs therefore leads to a larger saturation imbalance in this limit, which is seen in Fig.~\ref{fig:imbalance_L7} for seven sites. 

\section{Quantum dot coupled to a single reservoir at equilibrium}
\label{sec:equilibrium}

\begin{figure}[h!]
\includegraphics[width=0.4\linewidth]{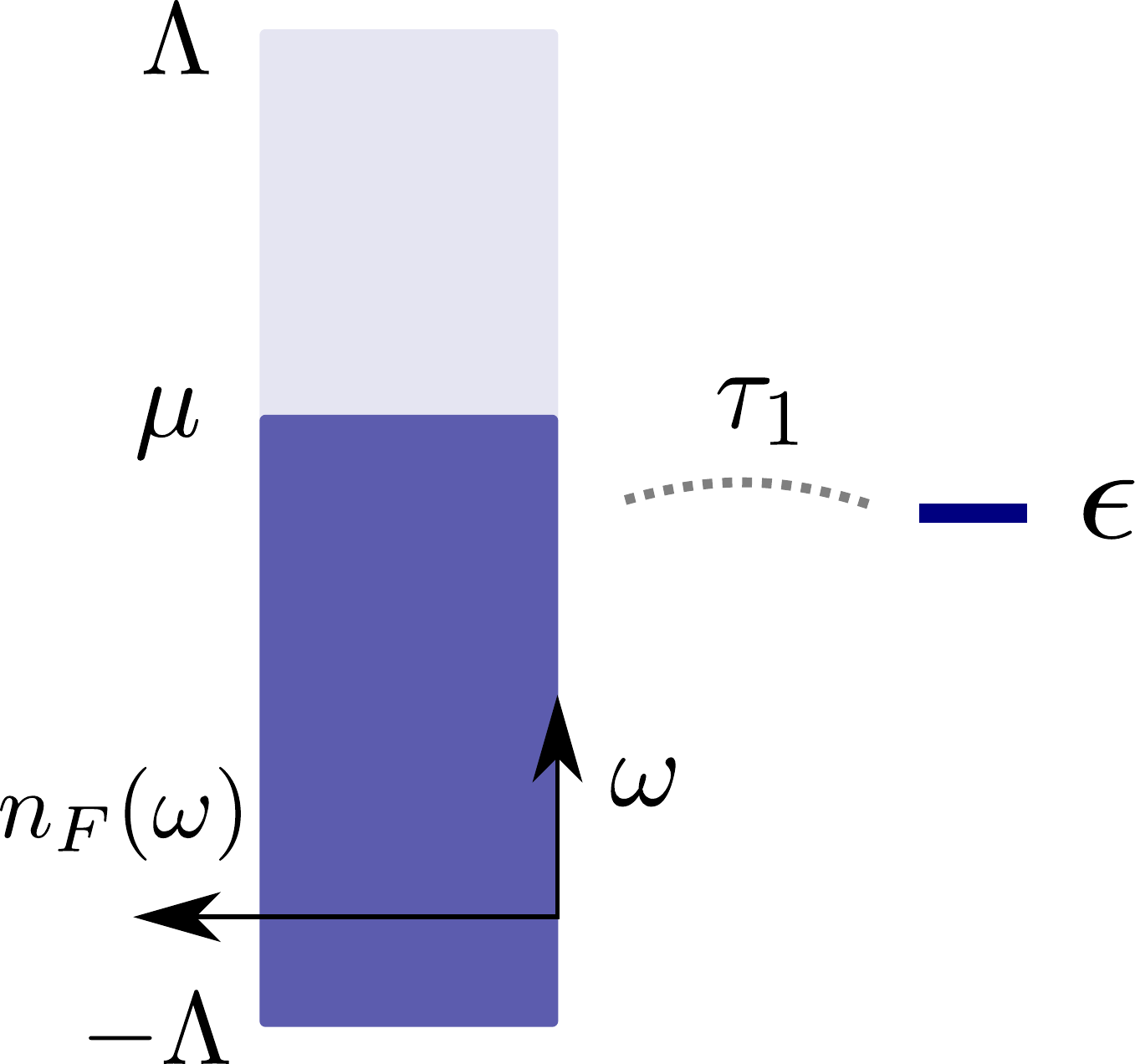}
\caption{ The energy diagram of a quantum dot with energy level $\epsilon$ coupled to a single reservoir. At zero temperature, states up to $\mu$ in the reservoir are filled and the rest are empty.}
\label{fig:energy_diagram}
\end{figure}

In this section, we introduce a simple model of a single quantum dot coupled to a reservoir at equilibrium, illustrated in Fig.~\ref{fig:energy_diagram}.
This model displays similar features in the occupation of the quantum dot as are found for the particle density and average density imbalance in a lattice coupled to reservoirs in the presence of the local loss. Namely, the broadening of the steps in the particle density imbalance with stronger coupling to the reservoirs, such as in Fig.~\ref{fig:imbalance_L7}, is also present in the single-dot equilibrium model. 
The connection exists only for the particle densities and not for currents since in the equilibrium system, there is no transport or particle loss. In the previous sections, we mostly discuss the limit of an unbounded spectrum of reservoir eigenvalues, but here we analyze in detail the effects of a finite cutoff in the spectrum. The simple model allows to distinguish the contribution of the continuous reservoir spectrum and that of discrete bound states to the quantum dot occupation, and to determine at which value of the cutoff the contribution of bound states is negligible.

We consider a linear dispersion relation of the reservoirs, so that the density of states is constant within the energy interval $[-\Lambda, \Lambda]$ indicated in Fig.~\ref{fig:energy_diagram}, $\rho(\omega) = \rho_0 \Theta(\Lambda - |\omega|)$. 
The density of states is therefore discontinuous at $\omega = \pm \Lambda$, and as derived in Appendix~\ref{app:equilibrium}, these discontinuities lead to the existence of two bound states at discrete energies outside the reservoir energy continuum. Spatially, the bound states have overlap with both the quantum dot and the reservoir. They therefore contribute to the occupation of the quantum dot.
While a constant density of states is the simplest choice and can be used as an approximation of more complex situations, it is exact for example for the quadratic dispersion relation in two dimensions. Discontinuities or singularities in the density of states lead to the occurrence of bound states also for example in the case of a quadratic dispersion relation in one dimension, or a one- or two-dimensional cosine dispersion.

\begin{figure}[h!]
\includegraphics[width=0.8\linewidth]{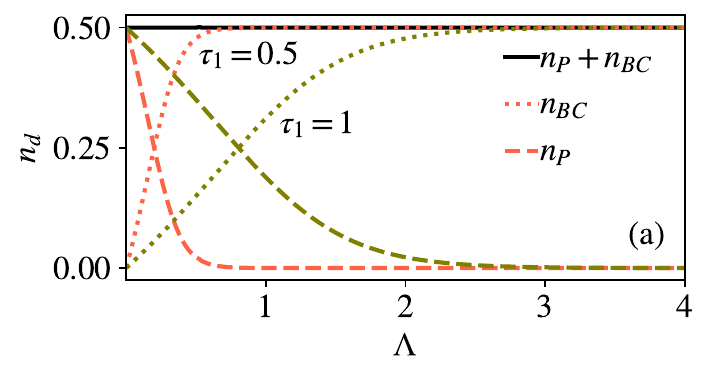}
\includegraphics[width=0.8\linewidth]{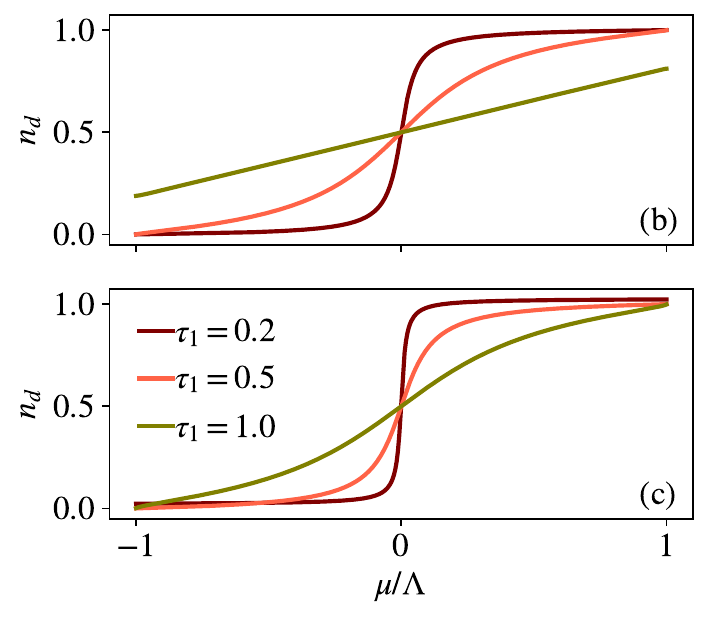}
\caption{(a) The two contributions to the quantum dot occupation, given by Eqs.~(\ref{eq:branch_cut}) and~(\ref{eq:poles}), indicated by the dotted and dashed lines. The different colors correspond to different values of $\tau_1$. The contribution of the bound states $n_P$ decays for increasing $\Lambda$ while the contribution of the reservoir energy continuum $n_{BC}$ increases. When $\mu = \epsilon = 0$, the occupation is 0.5 independent of the cutoff $\Lambda$. (b, c) The occupation $n_d$ as a function of the reservoir chemical potential $\mu$ with $\epsilon = 0$. The step-like change in the occupation is smoothened out for larger tunneling amplitudes, similar to the average density imbalance in Fig.~\ref{fig:imbalance_L7}. (b) For a small cutoff $\Lambda = 1$, the bound-state contribution is finite when the coupling is large ($\tau_1 = 1$). (c) For a larger cutoff $\Lambda = 3$, the bound-state contribution is negligible for all values of~$\tau_1$ shown here.}
\label{fig:single_dot}
\end{figure}

The equilibrium occupation of the quantum dot is given by $n_d = \int_{-\infty}^{\infty} d\omega n_F(\omega - \mu) A(\omega)$, where $A(\omega)$ is the spectral function at the quantum dot. As detailed in Appendix~\ref{app:equilibrium}, this integral has contributions arising both from the reservoir energy continuum and the bound states outside the continuum, corresponding to a branch cut and poles of the retarded Green's function on the real axis, respectively. We can separate these two contributions to the quantum dot occupation: $n_d = n_{BC} + n_P$, where
\begin{equation}
n_{BC} = \frac{1}{\pi} \int_{-\Lambda}^{\Lambda} d\omega \frac{n_F(\omega - \mu) \Gamma}{\left[ \omega - \epsilon - \Sigma_1(\omega) \right]^2 + \Gamma^2}
\label{eq:branch_cut}
\end{equation}
and
\begin{equation}
n_P = \sum_{E_b} \frac{n_F(E_b - \mu)}{\left| 1 - \partial_{\omega} \Sigma_1(\omega) \right|_{\omega = E_b}}.
\label{eq:poles}
\end{equation}
Here, $\Sigma_1(\omega) = \frac{\Gamma}{\pi} \left(\ln|\omega + \Lambda| - \ln|\omega - \Lambda| \right) $ is the real part of the retarded self-energy. The two bound states occur at frequencies $E_b$, which are solved from $E_b-\epsilon-\Sigma_1(E_b)=0$.
Figure~\ref{fig:single_dot}(a) shows how $n_{BC}$ and $n_P$ depend on the cutoff and the coupling to the reservoir in the symmetric situation where the quantum dot energy level $\epsilon$ and the reservoir chemical potential $\mu$ are both in the middle of the reservoir energy continuum, $\epsilon = \mu = 0$. While the bound-state contribution decays with increasing cutoff, the continuum contribution correspondingly increases so that they sum up to $n_d = 0.5$. A smaller coupling $\tau_1$ leads to a faster decay of $n_P$.

In Fig.~\ref{fig:single_dot}(b, c), we plot the occupation of the quantum dot as a function of the chemical potential $\mu$ of the reservoir, with a fixed value of the cutoff $\Lambda$. For $\mu = -\Lambda$, the occupation is given by the contribution of the bound state below the reservoir continuum. In panel~(b), we fix $\Lambda = 1$, for which this contribution is negligible for small couplings $\tau_1 = 0.2$ and $\tau_1 = 0.5$. The quantum dot is therefore empty when the chemical potential of the reservoir is equal to the lower cutoff. For $\tau_1 = 1$, the bound-state contribution is $n_P \approx 0.2$, so that the quantum dot is partly filled already at $\mu = -\Lambda$. On the other hand, for chemical potentials $\mu \leq \Lambda$, the dot is never fully filled since the finite contribution of the bound state above the reservoir continuum is not included.
For a larger cutoff, such as $\Lambda = 3$ in panel~(c), the bound-state contribution is negligible for all values of the coupling shown here, and the occupation grows from close zero to approximately one when the reservoir chemical potential changes from $\mu = - \Lambda$ to $\Lambda$. For a large coupling $\tau_1 = 1$, there is a smooth change in occupation, whereas for decreasing values of $\tau_1$, the change becomes step-like. This behavior is similar to the one observed for the density imbalance in larger lattices in Sec.~\ref{sec:imbalance}. 

Furthermore, the particle density at the outermost sites of a three-site lattice coupled to reservoirs at either end is given by equations similar to Eqs.~(\ref{eq:branch_cut}) and~(\ref{eq:poles}) in the $\gamma \to \infty$ limit (see Appendix~\ref{app:three-site_limit}). The contribution of the reservoir continuum is given by Eq.~(\ref{eq:branch_cut}), replacing $\mu$ by the chemical potential of either the left or right reservoir. In the bound-state contribution of Eq.~(\ref{eq:poles}), $\mu$ is replaced by $\epsilon$. The simple model of a quantum dot coupled to a single reservoir at equilibrium produces therefore an expression for the particle density which is almost identical to the nonequilibrium occupation in the three-site lattice in this limit.

\section{Conclusions}
\label{sec:conclusions}

In this paper, we characterize transport in the nonlinear regime and properties of nonequilibrium steady states in a lattice coupled to free-fermion reservoirs, subjected to a local particle loss at the center site. We find that the nonlinear current-voltage characteristics shows interesting step-like features. These steps are either smoothened out or preserved in the presence of the particle loss. Similar features appear in the loss current and the particle density imbalance between the left and right halves of the lattice. An explanation for these features and their modification by the dissipation is found through the single-particle eigenstates of an isolated lattice. We show that features arising from spatially symmetric eigenstates are smoothened out by the local dissipation, while those arising from antisymmetric eigenstates are more robust to dissipation or enhanced by it.

For nonzero voltages within the lattice energy band, the momentum distribution in the lattice shows two discontinuities at Fermi momenta corresponding to the chemical potential in either reservoir. In the absence of dissipation, transport in the lattice is ballistic and the momentum distribution is independent of position. This is connected to a uniform density distribution. A local particle loss depletes alternating momentum states depending on their spatial overlap with the lossy site, while the Fermi momentum is unchanged by the local dissipation. The preservation of the Fermi momentum is observed in the wavevector of Friedel oscillations in the density distribution, which in most cases is unchanged by the dissipation. We furthermore introduce an equilibrium model of a quantum dot coupled to a single reservoir. This simple model displays the same broadening characteristics with increasing coupling to the reservoirs as is observed in the average density imbalance of the nonequilibrium model in the presence of dissipation.

The nonequilibrium phenomena reported here are relevant for transport in mesoscopic wires~\cite{PothierDevoret1997}, where local electron losses could be implemented through additional leads~\cite{MorpurgoVanWees1998,BaselmansKlapwijk1999,MorpurgoKlapwijk2000,CrosserBirge2006}. In cold-atom experiments, transport and nonequilibrium steady-state properties have recently been explored in the presence of a local particle losses and lattice potentials~\cite{BarontiniOtt2013,LabouvieOtt2016,LebratEsslinger2018,CormanEsslinger2019,LebratEsslinger2019,BenaryOtt2022}. It would be interesting to compare the effects of a local particle loss to those of local dephasing~\cite{TonielliMarino2020,JinGiamarchi2022,WillFleischhauer2022}. Furthermore, the theoretical analysis applied here could also be used for studying transport through periodically driven (lossy) impurities~\cite{LudovicoSanchez2016,ReyesEggert2017,KamarGiamarchi2017,HuebnerSheikhan2022}, where resonance effects are expected to occur.

\begin{acknowledgments}

We thank C. Berthod for comments on the manuscript, and S. Diehl, T. Esslinger, P. Fabritius, M.-Z. Huang, J. Mohan, H. Ott, M. Talebi, S. Uchino, and S. Wili for helpful and inspiring discussions.
We acknowledge funding from the Deutsche Forschungsgemeinschaft (DFG, German Research Foundation) in particular under project number 277625399 - TRR 185 (B3) and project number 277146847 - CRC 1238 (C05) and under Germany’s Excellence Strategy – Cluster of Excellence Matter and Light for Quantum Computing (ML4Q) EXC 2004/1 – 390534769 and the European Research Council (ERC) under the Horizon 2020 research and innovation programme, grant agreement No.~648166 (Phonton). This work was supported in part by the Swiss National Science Foundation under Division II (Grant No.~2000020-188687). 
\end{acknowledgments}

\bibliographystyle{apsrev4-1-with-titles}
\bibliography{references_Bonn.bib}

\appendix

\section{Loss current, derivation}
\label{app:loss_current}

In this appendix, we show that in the considered setup, the loss current $I_{\text{loss}} = -\frac{d}{dt}\braket{N_L + N_R}$ is proportional to the particle number at the lossy site, $I_{\text{loss}} = \gamma \braket{n_0}$. In order to show this, we use Eq.~(\ref{eq:master_equation}),
\begin{align*}
\frac{d}{dt} \braket{n_0} &= \frac{d}{dt} \text{Tr} \left( n_0 \rho(t) \right) \\
&= -i \text{Tr} \left(n_0 [H, \rho] \right) \\
&\hspace{1cm} + \gamma \text{Tr} \left( n_0 d_0 \rho d_0^\dagger - \frac{1}{2} n_0 \{ d_0^\dagger d_0
, \rho \} \right) \\
&= -i \braket{ [n_0, H] } - \gamma \braket{n_0}.
\end{align*}
In the Hamiltonian part, only the tunneling terms contribute, giving
\begin{align}
\begin{split}
\braket{ [n_0, H] } &= -\tau \Big(\braket{d_0^\dagger d_{-1}} - \braket{d_{-1}^{\dagger} d_0} + \braket{d_0^\dagger d_1} \\
&\hspace{3cm}- \braket{d_1^{\dagger} d_0} \Big).
\end{split}
\label{eq:n0_commutator}
\end{align}
Similarly, we can obtain the equation of motion for the other sites $j\not = 0$
\begin{align*}
\frac{d}{dt} \braket{n_j} &= i \tau \Big(\braket{d_j^\dagger d_{j - 1}} - \braket{d_{j - 1}^{\dagger} d_j} + \braket{d_j^\dagger d_{j + 1}} - \braket{d_{j + 1}^{\dagger} d_j} \Big).
\end{align*}

In the steady state, the time-derivative vanishes for all lattice sites, i.e. $\frac{d}{dt} \braket{n_j} = 0$. Thus, in Eq.~(\ref{eq:n0_commutator}) we may replace
\begin{align*}
\braket{d_0^\dagger d_{-1}} - \braket{d_{-1}^{\dagger} d_0} &= \braket{d_{-1}^{\dagger} d_{-2}} - \braket{d_{-2}^\dagger d_{-1}} = \dots \\
&= \braket{d_{-l}^{\dagger} \psi_L(\mathbf{0})} - \braket{\psi_L^\dagger(\mathbf{0}) d_{-l}}
\end{align*}
and similarly
\begin{equation*}
\braket{d_0^\dagger d_1} - \braket{d_1^{\dagger} d_0} = \braket{d_l^{\dagger} \psi_R(\mathbf{0})} - \braket{\psi_R^\dagger(\mathbf{0}) d_l}.
\end{equation*}
The time derivative of $\braket{n_0}$ can then be written as
\begin{align*}
\frac{d}{dt} \braket{n_0} &= -\frac{d}{dt} \braket{N_L + N_R} - \gamma \braket{n_0} = 0
\end{align*}
which leads to the relation $I_{\text{loss}} = \gamma \braket{n_0}$.

\section{Reservoirs with a finite energy continuum}

\subsection{General expressions}
The local reservoir Green's function of Eq.~(\ref{eq:greens_function_r0}) can generally be written as a sum of its real and imaginary parts, $G^{\mathcal{R}, \mathcal{A}}_{L/R}(\mathbf{r} = \mathbf{0}, \omega) = A(\omega) \mp i B(\omega)$ (see also Appendix~\ref{app:equilibrium}). For reservoirs with a constant density of states, we have the real part $A(\omega) = \rho_0 \left(\ln|\omega + \Lambda| - \ln|\omega - \Lambda| \right)$,
and the imaginary part is $B(\omega) = \pi\rho_0 \Theta(\Lambda - |\omega|)$. In the limit $\Lambda \to \infty$, the real part vanishes and the imaginary part is the constant $\pi \rho_0$. When the cutoff $\Lambda$ is finite, the imaginary part $B(\omega)$ is zero for $|\omega| > \Lambda$, and to evaluate correctly the integrals of Eq.~(\ref{eq:conserved_current_integral}) and~(\ref{eq:occupation_integral}), one has to take into account the infinitesimal imaginary term $i\eta$ at $|\omega| > \Lambda$. We therefore keep $i\eta$ in the local Green's functions for the lattice site $j = 0$ with the particle loss: $G_{j = 0}^{\mathcal{R}, \mathcal{A}} = \left( \omega - \epsilon \pm i\gamma/2 \pm i\eta\right)^{-1}$ and $[G_{j = 0}^{-1}]^{\mathcal{K}} = i \gamma + 2 i \eta [1 - 2n_F(\omega - \epsilon)]$. The expression for the conserved current is now
\begin{equation}
I = \int_{-\Lambda}^{\Lambda} \frac{d \omega}{2 \pi} \tilde{g}(\omega) \left[ n_L(\omega) - n_R(\omega) \right],
\label{eq:conserved_current_finite_cutoff}
\end{equation}
where the modified function $\tilde{g}(\omega)$ contains the finite real part of the local reservoir Green's function. It is proportional to the imaginary part and therefore zero for $|\omega| > \Lambda$. The loss current is
\begin{align}
\begin{split}
I_{\text{loss}} &= \gamma \int_{-\Lambda}^{\Lambda} \frac{d \omega}{2 \pi} \tilde{f}(\omega) \left[ n_L(\omega) + n_R(\omega) \right] \\
&\hspace{1cm}+\gamma \int_{-\infty}^{-\Lambda} \frac{d \omega}{2 \pi} f_{\eta}(\omega) n_F(\omega - \epsilon) \\
&\hspace{1cm}+\gamma \int_{\Lambda}^{\infty} \frac{d \omega}{2 \pi} f_{\eta}(\omega) n_F(\omega - \epsilon).
\end{split}
\label{eq:loss_current_finite_cutoff}
\end{align}
The Fermi function $n_F(\omega - \epsilon)$ appears in the integrals over $|\omega| > \Lambda$, and $f_{\eta}(\omega)$ is a function that depends on the infinitesimal  $i\eta$. In the limit $\Lambda \to \infty$, we have $\tilde{g}(\omega) \to g(\omega)$ and $\tilde{f}(\omega) \to f(\omega)$ and the second and third line in Eq.~(\ref{eq:loss_current_finite_cutoff}) vanish.

\subsection{Conserved and loss current for the quantum dot}
\label{app:finite_continuum}

In this section, we discuss a single lossy quantum dot coupled to leads. When the cutoff $\Lambda$ is finite, the integrand $\tilde{g}(\omega)$ in the expression~(\ref{eq:conserved_current_finite_cutoff}) for the conserved current has the form
\begin{equation}
\tilde{g}(\omega) = \frac{4 \Gamma (\gamma + 4 \Gamma)}{(\gamma + 4 \Gamma)^2 + 4 \left(\omega - \epsilon - 2 \tau_1^2 A(\omega) \right)^2}.
\label{eq:quantum_dot_integrand}
\end{equation}
Compared to the limit of infinite reservoirs taken in Eq.~(\ref{eq:quantum_dot_integrand}), a finite cutoff and the presence of a finite $A(\omega)$ term leads to a shift of the maximum of the Lorentzian distribution. The position of the maximum approaches $\omega = \epsilon$ in the limit $\tau_1 \to 0$. 
For unbounded reservoirs, as in Fig.~\ref{fig:conserved_current_qd}, the current saturates at $I = \Gamma$ in the limit of infinite voltage, independently of the loss rate $\gamma$.
Keeping a finite cutoff $\Lambda$ instead leads to a decay of the saturation value with increasing loss as $\sim 1/\gamma$.

\begin{figure}[h!]
\includegraphics[width=\linewidth]{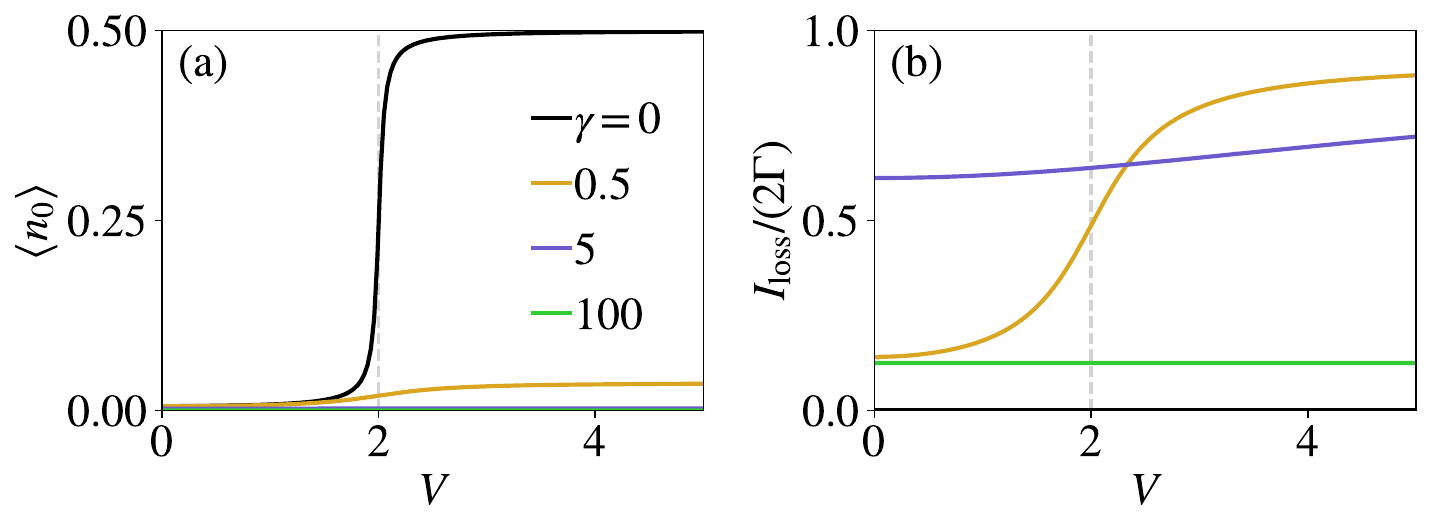}
\caption{(a) The quantum dot occupation and (b) the loss current as functions of voltage for a finite cutoff $\Lambda = 10$. When $\Lambda$ is finite, the loss current has a nonmonotonic dependence on $\gamma$. The coupling to the reservoirs is $\tau_1 = 0.1$ and $\epsilon = 1$.}
\label{fig:loss_current_finite_cutoff}
\end{figure}

For a single quantum dot coupled to reservoirs, the function $\tilde{f}(\omega)$ in Eq.~(\ref{eq:loss_current_finite_cutoff}) is
\begin{equation}
\tilde{f}(\omega) = \frac{8 \Gamma}{\left(\gamma + 4 \Gamma \right)^2 + 4 \left[ \omega - \epsilon - 2 \tau_1^2 A(\omega) \right]^2}
\label{eq:loss_current_f}
\end{equation}
and 
\begin{equation}
f_{\eta}(\omega) = \frac{8 \eta}{\left(\gamma + 2\eta \right)^2 + 4 \left[ \omega - \epsilon - 2 \tau_1^2 A(\omega) \right]^2}.
\label{eq:f_eta}
\end{equation}
Figure~\ref{fig:loss_current} shows that the dependence of the loss current on dissipation is very different when there is a finite cutoff compared to an unbounded reservoir spectrum. At zero voltage, an unbounded spectrum leads to saturation of the loss current at $2\Gamma$ when $\gamma \to \infty$. This is also seen at nonzero voltages in Fig.~\ref{fig:loss_current_qd}. Here, we show the dependence of the loss current on voltage when the cutoff is finite. In this case, it is given by Eqs.~(\ref{eq:loss_current_finite_cutoff}), (\ref{eq:loss_current_f}), and~(\ref{eq:f_eta}). At weak dissipation, the loss current has a step-like behavior, as seen in Fig.~\ref{fig:loss_current_finite_cutoff}(b), while at larger $\gamma$, the step is smoothened out. The loss current approaches zero at all voltages when $\gamma \to \infty$. The loss current therefore has a nonmonotonic dependence on the loss rate $\gamma$ at all voltages.

\subsection{Voltage-independent loss current at $\epsilon = 0$}
\label{app:loss_current_voltage}

When the function $\tilde{f}(\omega)$ in Eq.~(\ref{eq:loss_current_f}) is an even function of $\omega$, the loss current is independent of voltage. We show here explicitly that this occurs for the quantum dot and the three-site system when the momentum cutoff in the reservoirs is chosen symmetrically around zero, $k \in [-\Lambda/v_F, \Lambda/v_F]$, and the chemical potentials in the reservoirs are chosen symmetrically around this zero level, $\mu_{L, R} = \pm V/2$. In the expression for the loss current, Eq.~(\ref{eq:loss_current_finite_cutoff}), the function $\tilde{f}(\omega)$ is given by Eq.~(\ref{eq:loss_current_f}) for a single quantum dot coupled to reservoirs.
For three sites coupled to reservoirs, the function $\tilde{f}(\omega)$ is
\begin{widetext}
\begin{align*}
\tilde{f}(\omega) = \frac{8 \Gamma \tau^2}{4 \left[ 2 \tau^2 + \omega \left( A(\omega) \tau_1^2 - \omega \right) \right]^2 + \left(\Gamma \gamma + 4 \tau^2 \right)^2 + \gamma^2\left( A(\omega) \tau_1^2 - \omega) \right)^2 - 16 \tau^4 + 4 \omega^2}.
\end{align*}
For both the quantum dot and three sites, the function $f_{\eta}(\omega)$ is independent of voltage, and at $\epsilon = 0$, $\tilde{f}(\omega)$ is an even function of $\omega$ in both cases. We can rewrite 
\begin{align*}
\int_{-\Lambda}^{\Lambda} &\frac{d \omega}{2 \pi} \tilde{f}(\omega) \left[ n_L(\omega) + n_R(\omega) \right] \\
&= \int_0^{\Lambda} \frac{d \omega}{2 \pi} \tilde{f}(\omega) \left[ n_F\left( \omega - \frac{V}{2} \right) + n_F\left( \omega + \frac{V}{2} \right) \right] 
+ \int_0^{\Lambda} \frac{d \omega}{2 \pi} \tilde{f}(-\omega) \left[ n_F\left( -\omega - \frac{V}{2} \right) + n_F\left( -\omega + \frac{V}{2} \right) \right] \\
&= \int_0^{\Lambda} \frac{d \omega}{2 \pi} \tilde{f}(\omega) \Big[ n_F\left( \omega - \frac{V}{2} \right) + n_F\left( -\omega + \frac{V}{2} \right) 
+ n_F\left( \omega + \frac{V}{2} \right) + n_F\left( -\omega - \frac{V}{2} \right) \Big] 
= 2 \int_0^{\Lambda} \frac{d \omega}{2 \pi} \tilde{f}(\omega).
\end{align*}
\end{widetext}
Here, we have used $n_F(\omega) + n_F(-\omega) = 1$. In the symmetric situation $\epsilon = 0$, the loss current is therefore independent of voltage.

\section{Density imbalance in the three-site system}
\label{app:three_sites}

The saturation value of the particle number imbalance in Fig.~\ref{fig:imbalance_L51} is reproduced by the simple function of Eq.~(\ref{eq:imbalance_expansion}). This function is motivated by taking the $V \to \infty$ limit of the imbalance in the three-site model and expanding the result at large $\gamma$. The density imbalance between the left- and rightmost sites is given by
\begin{equation}
\delta n = \int_{-\infty}^{\infty} \frac{d\omega}{2\pi} h(\omega) \left[ n_L(\omega) - n_R(\omega) \right],
\end{equation}
where
\begin{align*}
h(\omega) &= \frac{2 \Gamma}{\Gamma^2 + \omega^2} \\
&\hspace{1mm}\times \frac{4 \gamma \Gamma \tau^2 + \gamma^2(\Gamma^2 + \omega^2) + 4 \omega^2(\Gamma^2 - 2\tau^2 + \omega^2)}{(\gamma\Gamma + 4\tau^2)^2 + \left[ \gamma^2 + 4(\Gamma^2 - 4\tau^2) \right] \omega^2 + 4 \omega^4}.
\end{align*}
Here, we have set $\epsilon = 0$. To find the saturation value at infinite voltage, we set $n_L(\omega) \to 1$ and $n_R(\omega) \to 0$. We furthermore expand the integrand $h(\omega)$ in the vicinity of $\gamma \to \infty$. The terms up to the first order in $1/\gamma$ integrate to 
\begin{equation}
1 - 2\tau^2/(\gamma\Gamma),
\label{eq:first_order}
\end{equation}
which coincides with the $\gamma \to \infty$ expansion of the function~(\ref{eq:imbalance_expansion}). Equation~(\ref{eq:imbalance_expansion}) reproduces the exact results for $M = 3$ and $M = 51$ better than the first-order expansion~(\ref{eq:first_order}) since it is zero at $\gamma = 0$.

\section{Quantum dot coupled to a single reservoir at equilibrium}
\label{app:equilibrium}

A quantum dot coupled to a single reservoir at equilibrium displays similar features in the occupation of the quantum dot as are found for the particle density in a lattice coupled to reservoirs in the limit of infinitely strong dissipation. In this limit, the lattice is effectively cut in half at the lossy site. We analyze in detail the effects of a finite cutoff in the reservoir spectrum, which leads to the appearance of bound states with discrete energies outside the reservoir energy continuum. We also derive Eqs.~(\ref{eq:branch_cut}) and~(\ref{eq:poles}).

\subsection{Occupation of the quantum dot}

The occupation of the quantum dot can be written in terms of the spectral function at the quantum dot $A(\omega)$~\cite{Mahan2000}, 
\begin{equation}
n_d = \int_{-\infty}^{\infty} d\omega n_F(\omega - \mu) A(\omega).
\end{equation}
The Fermi function $n_F(\omega - \mu)$ takes into account the occupation of energy levels in the reservoir.
The spectral function is related to the retarded Green's function of the quantum dot $G_d(\omega)$,
\begin{equation}
A(\omega) = -\frac{1}{\pi} \text{Im} \; G_d(\omega),
\label{eq:spectral_function}
\end{equation}
where
\begin{equation}
G_d(\omega) = \frac{1}{\omega - \epsilon - \Sigma(\omega) + i \eta}.
\label{eq:retarded_greens_function}
\end{equation} 
Physically, $A(\omega)$ gives the excitation spectrum of the quantum dot, which for an isolated dot would be a delta function at $\omega = \epsilon$ but due to the coupling to the reservoir has a finite width. This width is connected to a finite lifetime of particles at the quantum dot, and is determined by the imaginary part of the retarded self-energy $\Sigma(\omega)$ in Eq.~(\ref{eq:retarded_greens_function}). Points in the spectrum where the imaginary part is zero and the real part finite correspond to bound states with an infinite lifetime, as discussed below.

We find the quantum dot occupation as 
\begin{align}
\begin{split}
n_d = -\frac{1}{\pi} &\int_{-\infty}^{\infty} d \omega n_F(\omega - \mu) \\
&\times \text{Im} \left[\frac{1}{\omega - \epsilon - \Sigma_1(\omega) - i \Sigma_2(\omega) + i\eta} \right],
\end{split}
\label{eq:single_dot_occupation}
\end{align}
where the retarded self-energy is written as a sum of the real and imaginary parts, $\Sigma(\omega) = \Sigma_1(\omega) + i \Sigma_2(\omega)$. 
The imaginary part $\Sigma_2(\omega)$ is responsible for a branch cut on the real axis in the interval $\omega \in [-\Lambda, \Lambda]$, where $\Sigma_2(\omega)$ is nonzero. Additionally, $G_d$ can have poles on the real axis when $\omega - \epsilon - \Sigma_1(\omega) = 0$.
We can then write Eq.~(\ref{eq:single_dot_occupation}) as $n_d = n_{BC} + n_P$, where
the term $n_{BC}$ is the contribution of the branch cut,
\begin{equation}
n_{BC} = \frac{1}{\pi} \int_{-\Lambda}^{\Lambda} d\omega \frac{n_F(\omega - \mu) \Sigma_2(\omega)}{\left[ \omega - \epsilon - \Sigma_1(\omega) \right]^2 + \left[ \Sigma_2(\omega) \right]^2}.
\label{eq:branch_cut_single_dot}
\end{equation}
The second term $n_P$ is the contribution of the poles, which arises from the part of the integral with $|\omega| > \Lambda$. Here, $\Sigma_2(\omega) = 0$ and
\begin{equation}
-\frac{1}{\pi} \text{Im} \; G_d(\omega) = \frac{1}{\pi} \frac{\eta}{\left[ \omega - \epsilon - \Sigma_1(\omega) \right]^2 + \eta^2}.
\label{eq:lorentzian_delta_limit}
\end{equation}
As $\eta$ is infinitesimal, this term is a delta function which gives the contribution of the poles,
\begin{align}
\nonumber
n_P &= \int_{-\infty}^{-\Lambda} d\omega \: n_F(\omega - \mu) \delta(\omega - \epsilon - \Sigma_1(\omega)) \\
\nonumber
&+ \int_{\Lambda}^{\infty} d\omega \: n_F(\omega - \mu) \delta(\omega - \epsilon - \Sigma_1(\omega)) \\
&= \sum_{E_b} \frac{n_F(E_b - \mu)}{\left| 1 - \partial_{\omega} \Sigma_1(\omega) \right|_{\omega = E_b}}.
\label{eq:bound_state_contribution}
\end{align}
Bound states occur at frequencies $E_b$, which are solved from
\begin{equation}
E_b - \epsilon - \Sigma_1(E_b) = 0.
\label{eq:bound_state_energies_general}
\end{equation}

\subsection{Retarded self-energy}

The retarded self-energy of the quantum dot is
\begin{align}
\Sigma(\omega) = \frac{\tau_1^2}{\mathcal{V}} &\sum_{k = -\Lambda/v_F}^{\Lambda/v_F} \frac{1}{\omega - E_k + i \eta} \\
&\to \tau_1^2 \rho_0 \int_{-\Lambda}^{\Lambda} \frac{dE}{\omega - E + i \eta},
\end{align}
where $\tau_1$ denotes the coupling between the quantum dot and the reservoir, and the sum is over the reservoir eigenmodes $k$. 
To evaluate the real and imaginary parts, we replace the sum over eigenmodes by an integral over energy, multiplied by the constant density of states $\rho_0$. We can rewrite the integral as 
\begin{align*}
\Sigma(\omega) = \tau_1^2 \rho_0 &\int_{-\Lambda}^{\Lambda} dE \Big[\mathcal{P} \left( \frac{1}{\omega - E} \right) - i \pi \delta(\omega - E) \Big] 
\nonumber \\
&= \Sigma_1(\omega) + i \Sigma_2(\omega),
\end{align*}
where 
\begin{equation}
\Sigma_1(\omega) = \frac{\Gamma}{\pi} \left(\ln|\omega + \Lambda| - \ln|\omega - \Lambda| \right) 
\label{eq:real_part}
\end{equation}
and $\Sigma_2(\omega) = -\Gamma \Theta(\Lambda - |\omega|)$
with $\Gamma = \pi \rho_0 \tau_1^2$.

\subsection{Bound states}
\label{app:bound_states}

\begin{figure}[h!]
\includegraphics[width=0.7\linewidth]{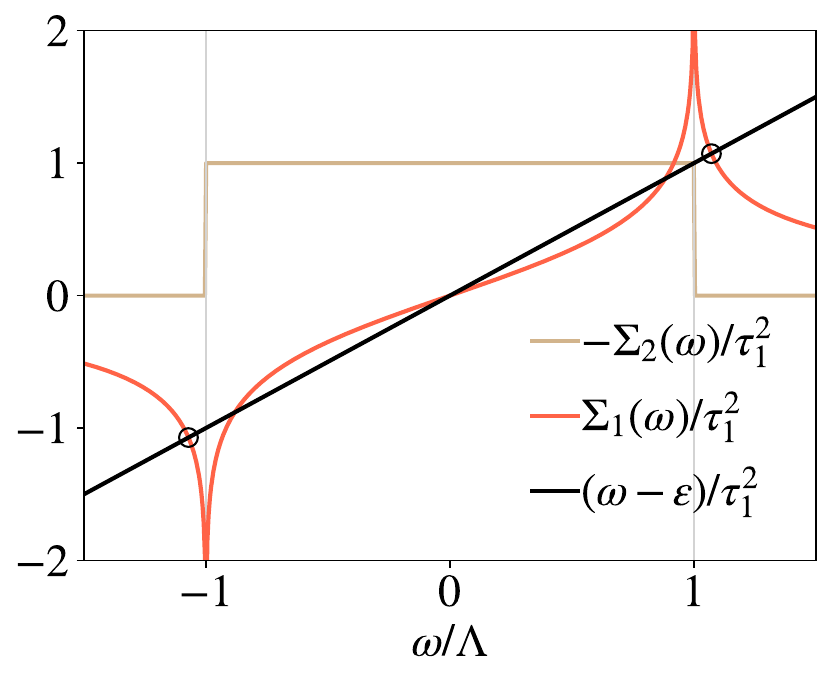}
\caption{The bound-state energies are found at the intersections marked by circles, as indicated by Eq.~(\ref{eq:bound_state_energies_general}). The $y$ axis is in units of $1/(\pi \rho_0)$.}
\label{fig:intersections}
\end{figure}

We substitute Eq.~(\ref{eq:real_part}) into Eq.~(\ref{eq:bound_state_energies_general}) to find the bound-state energies,
\begin{equation}
E_b - \epsilon - \frac{\Gamma}{\pi} \ln \left( \frac{E_b + \Lambda}{E_b - \Lambda} \right)  = 0,
\label{eq:bound_state_energies}
\end{equation}
where we have used $|E_b| > \Lambda$. The bound-state contribution to the integral is obtained as
\begin{align}
n_P = \sum_{E_b} \frac{1}{\left| 1 - \frac{\Gamma}{\pi} \left( \frac{1}{E_b + \Lambda} - \frac{1}{E_b - \Lambda} \right) \right|}.
\label{eq:bound-state_weight}
\end{align}
The bound-state contribution therefore depends on the value of the cutoff~$\Lambda$. Figure~\ref{fig:intersections}(a) shows the graphical solution of the bound-state energies from Eq.~(\ref{eq:bound_state_energies}): they are given by the values of $\omega$ where the line $\omega - \epsilon$ intersects with the real part of the self-energy $\Sigma_1(\omega)$ in the region $|\omega| > \Lambda$ where the imaginary part $\Sigma_2(\omega)$ is zero. From these bound-state energies, we calculate the bound-state weight given by Eq.~(\ref{eq:bound-state_weight}). Figure~(\ref{fig:intersections}) shows that this contribution decays rapidly with an increasing cutoff, while the contribution of the reservoir continuum increases. To solve the particle densities numerically for the 51-site lattice discussed in Sections~\ref{sec:friedel_oscillations} and~\ref{sec:imbalance}, we use in practice a finite cutoff. We choose a value of $\Lambda$ which is sufficient to consider only the contribution of the continuum. This means that to compute the occupation $n_d = \int_{-\infty}^{\infty} \frac{d \omega}{2 \pi} n_d(\omega)$, we limit the integration to the interval $\omega \in [-\Lambda, \Lambda]$.

\subsection{Three-site lattice in the $\gamma \to \infty$ limit}
\label{app:three-site_limit}

The occupation probability of the outermost sites in a three-site chain coupled to reservoirs at either end can be related to the equilibrium occupation of the single quantum dot in the limit of infinite particle loss $\gamma \to \infty$. In this limit, we obtain the expression $n_{\pm 1} = n_{\pm 1}^{BC} + n_{\pm 1}^P$, where the contribution of the reservoir continuum is
\begin{align}
n_{\pm 1}^{BC} =
&\int_{-\Lambda}^{\Lambda} \frac{d \omega}{2 \pi} \frac{2 \Gamma n_F(\omega - \mu_{L, R})}{\left[ \omega - \epsilon - \Sigma_1(\omega) \right]^2 + \Gamma^2},
\end{align}
equivalent to Eq.~(\ref{eq:branch_cut_single_dot}), and the contribution of the poles is obtained as
\begin{align}
n_{\pm 1}^P =
&\int_{-\infty}^{-\Lambda} \frac{d \omega}{2 \pi} \frac{2 \eta \: n_F(\omega - \epsilon)}{\eta^2 + \left[ \omega - \epsilon - \Sigma_1(\omega) \right]^2}
\label{eq:infinitesimal1} \\ 
&+ \int_{\Lambda}^{\infty} \frac{d \omega}{2 \pi} \frac{2 \eta \: n_F(\omega - \epsilon)}{\eta^2 + \left[ \omega - \epsilon - \Sigma_1(\omega) \right]^2}.
\label{eq:infinitesimal2}
\end{align}
We take the limit
\begin{equation}
\lim_{\eta \to 0} \frac{1}{\pi}\frac{\eta}{\eta^2 + \left[ \omega - \epsilon - \Sigma_1(\omega) \right]^2} = \delta \left( \omega - \epsilon - \Sigma_1(\omega) \right)
\end{equation}
to obtain
\begin{equation}
n_{\pm 1}^P = \sum_{E_b} \frac{n_F(E_b - \epsilon)}{\left| 1 - \partial_{\omega} \Sigma_1(\omega) \right|_{\omega = E_b}}.
\label{eq:bound_states_three_sites}
\end{equation}
This factor is equivalent to Eq.~(\ref{eq:bound_state_contribution}), apart from the argument of the Fermi distribution, which is $\omega - \epsilon$ instead of $\omega - \mu_L$ as in the equilibrium model. At zero temperature, when $\mu, \epsilon \in [-\Lambda, \Lambda]$, the two are equal.

\end{document}